\documentclass[aps,tightenlines,showpacs,amsmath,amssymb]{revtex4}%
\usepackage{graphicx}
\usepackage{epsfig}
\usepackage{graphics,color}
\usepackage{amssymb}
\usepackage[hypertex]{hyperref}
\usepackage{amsbsy}
\usepackage{bm}
\newcommand{\be}{\begin{equation}}
\newcommand{\ee}{\end{equation}}
\newcommand{\bse}{\begin{subequations}}
\newcommand{\ese}{\end{subequations}}
\newcommand{\bea}{\begin{eqnarray}}
\newcommand{\eea}{\end{eqnarray}}

\newcommand{\bean}{\begin{eqnarray*}}
\newcommand{\eean}{\end{eqnarray*}}

\begin{document}
\title{Phase diagram and thermodynamics of the Polyakov linear sigma model with three quark flavors}
\author{Hong Mao$^{1,2,3}$}
\email {maohong@hznu.edu.cn}
\author{Jinshuang Jin$^{2}$}
\author{Mei Huang$^{1,3}$}
\email {huangm@mail.ihep.ac.cn}
\address{1. Institute of High Energy Physics, Chinese Academy of Sciences, Beijing
100049, China\\
2. Department of Physics, Hangzhou Normal
University, Hangzhou 310036, China \\
3. Theoretical Physics Center for Science Facilities, Chinese Academy of Sciences, Beijing 100049, China}


\begin{abstract}
The phase diagram at finite temperature and density is investigated
in the framework of the Polyakov linear sigma model (PLSM) with
three light quark flavors in the mean field approximation. It is
found that in the PLSM, the three phase transitions, i.e, the chiral
restoration of $u,d$ quarks, the chiral restoration of $s$ quark,
and the deconfinement phase transition are independent. There exists
two-flavor quarkyonic phase at low density and three-flavor
quarkyonic phase at high density. The critical end point (CEP) which
separating the crossover from the first-order line in the PLSM model
is located at$(T_E, \mu_E)=(188~\mathrm{MeV},139.5~\mathrm{MeV})$.
In the transition region the thermodynamic properties and bulk viscosity over
entropy density ratio $\zeta/s$ are also discussed in the PLSM.
\end{abstract}


\maketitle

\section{Introduction}

It is widely believed that at sufficiently high temperatures and
densities there is a quantum chromodynamics (QCD) phase transition
between normal nuclear matter and quark-gluon plasma (QGP), where
quarks and gluons are no longer confined in
hadrons\cite{Rischke:2003mt}. The theoretical and experimental
investigation of QGP is one of the most challenging problems in high
energy physics. The main target of heavy ion experiments at the
Relativistic Heavy Ion collider (RHIC), the forthcoming Large Hadron
Collider (LHC) and FAIR at GSI is to create such form of matter and
study its properties.

There are two different phase transition which are associated with
two opposite quark mass limits. For $N_f$ massless quark flavors,
the QCD Lagrangian possesses a chiral $U(N_f)_R\times
U(N_f)_L=SU(N_f)_R\times SU(N_f)_L\times U(1)_V\times U(1)_A$
symmetry, here $V=R+L$, while $A=R-L$. However this symmetry does
not appear in the low energy particle spectrum, it is spontaneously
broken to the diagonal $SU(N_f)_V$ group of vector transformation by
a non-vanishing expectation value of the quark-antiquark condensate,
$\langle \overline{q}_Rq_L\rangle\neq 0$. This process involves
$N^2_f $ Goldstone bosons which dominate the low-energy dynamics of
the theory. The $U(1)_V$ symmetry is always respected and thus plays
no role in the symmetry breaking pattern considered in the following
discussion. The axial $U(1)_A$ symmetry is broken to $Z(N_f)_A$ by
instanton effects \cite{'tHooft:1976up}\cite{'tHooft:1976fv}.
Consequently, one of the $N^2_f$ Goldstone bosons becomes massive,
leaving $N^2_f-1$ Goldstone bosons. The $SU(N_f)_R\times
SU(N_f)_L\times U(1)_A$ group is also explicitly broken by the
effects of nonzero quark masses. The $N^2_f-1$ low energy degrees of
freedom then become pseudo-Goldstone bosons. For $M\leq N_f$
degenerate quark flavors, an $SU(M)_V$ symmetry is
preserved\cite{Roder:2003uz}\cite{Lenaghan:2000ey}. It is generally
believed that at sufficiently high temperature and density there
should be a transition from ordinary hadronic matter to a chirally
symmetric plasma of quark and gluons. At temperature of about 170
MeV, Lattice QCD calculations indicate that this symmetry is
restored\cite{Karsch:2001cy}. The order of the phase transition
seems to depend on the mass of the non-strange $u$ and $d$ quarks,
and the mass of the strange quark $m_s$, and at the temperature on
the order of 200 MeV, heavier quark flavors do not play an essential
role.

On the contrary, in the heavy quark limit, QCD reduces to a pure
$SU(N_c)$ gauge theory which is invariant under a global $Z(N_c)$
center symmetry, where the Polyakov loop serves as an order
parameter\cite{Polyakov:1978vu, Susskind:1979up, Svetitsky:1982gs,
Svetitsky:1985ye}. In general, the Polyakov loop is a complex scalar
field and it is related to the free energy of a static quark in the
gluon medium, it vanishes in the confining phase means that the
quark has infinite free energy and takes a finite value in the
deconfined phase. It is linked to the $Z(N_c)$ center symmetry of
the $SU(N_c)$ gauge theory, thus the confining phase is center
symmetric, while the center symmetry is spontaneously broken in the
deconfined phase. In the present of dynamical quarks the center
symmetry is explicitly broken. No order parameter is established for
the deconfinement transition in this case, but the Polyakov loop
still serves as an indicator of a rapid crossover towards
deconfinement\cite{Fukushima:2002bk}. On the other hand, the chiral
transition has a well-defined order parameter in the chiral limit of
massless quarks: the quark-antiquark condensate, $\langle
\overline{q}_Rq_L\rangle\neq 0$, which and its dynamical generation,
is the basic element of the linear sigma model and the
Nambu-Jona-Lasinio model.

Recently, the effective chiral models for QCD with $U(N_f)_R\times
U(N_f)_L$ symmetry, such as the Nambu-Jona-Lasinio model(NJL) or,
via bosonization, the linear sigma model (LSM) could be combined
with the Polyakov loop which allows to investigate both, the chiral
and the deconfinement phase transition. These models named as the
Polyakov-Nambu-Jona-Lasinio model (the PNJL) \cite{Fukushima:2003fw,
Ratti:2005jh,Fukushima:2008wg}, and the Polyakov linear sigma model
(PLSM) or the Polyakov quark meson model (PQM)\cite{Schaefer:2007pw,
Kahara:2008yg, Schaefer:2008ax}, are widely studied in recent years.
In the PNJL and PLSM model, where quarks develop quasiparticle
masses by propagating in the chiral condensate, while they couple at
the same time to a homogeneous background (temporal) gauge field
representing the Polyakov loop dynamics. The full QCD thermodynamics
at zero and finite quark chemical potential have been investigated
in Refs.\cite{Schaefer:2007pw,
Kahara:2008yg}\cite{Megias:2004hj,Ghosh:2006qh,Ratti:2006wg,
Mukherjee:2006hq, Roessner:2006xn, Zhang:2006gu, Sasaki:2006ww,
Ratti:2007jf, Fu:2007xc,
Ciminale:2007sr,Abuki:2008nm,GomezDumm:2008sk, Fu:2009zs}. However,
most studies focus on the PNJL model with two and three quark
flavors. As we know both of the NJL model and the LSM model for the
phenomenology of QCD can be parameterized to describe equally well
the vacuum structure at $T=\mu=0$ MeV, but these models treat the
contribution of the Dirac sea differently, in the NJL model it is
included explicitly up to a momentum cutoff $\Lambda$, while in the
sigma model this contribution is renormalized out. So that the
extended PLSM model which is augmented with the Polyakov loop has
the benefit of renormalizability. In this paper, calculations for
the thermodynamic potential and the phase diagram will be performed
at the mean-field level as that of the PNJL model, however,
eventually, such model will be investigated beyond mean-field by the
loop expansion \cite{Herpay:2006vc}, especially by using the CJT
effective potential\cite{Cornwall:1974vz}.

In present paper, we extend the Polyakov linear sigma model with two
quark flavors \cite{Schaefer:2007pw, Kahara:2008yg} to three quark
flavor\cite{Schaefer:2008ax} to investigate the phase diagram at
finite chemical potential. Moreover, since the study of the equation
of state in QCD with the $2+1$ quark flavors by the lattice QCD
simulations with almost physical quark masses have recently been
reported in Ref\cite{Cheng:2007jq} at zero chemical potential but
finite temperature, it is interesting to describe QCD thermodynamics
in the PLSM with three quark flavors, and compare with the recent
results of the lattice simulations at zero chemical potential,
especially, the essentials of QCD thermodynamics around the critical
temperature $T_c$. The outline of the paper is as follows: in the
next section we introduce the PLSM model with three quark flavors,
some symmetry breaking patterns in the vacuum are briefly discussed,
and parameters are fixed. In Sect. III, after obtaining the
effective potential in the mean field approximation, we explore the
phase diagram at $T-\mu$ plane. Section IV is devoted to derive the
thermodynamic properties of the system at zero chemical potential,
i.e. the pressure, the equation of state, the square of the speed of
sound, the specific heat, the trace anomaly of the energy-momentum
tensor and the bulk viscosity, all these thermodynamical observables
are compared with the lattice QCD data. At the end, we give
discussions and summary in Sec.V.

\section{The Model}

Following Ref.\cite{Schaefer:2007pw}, we introduce a generalized
Lagrangian of the linear sigma model for $N_f=3$ quarks and $N_c=3$
color degrees with quarks coupled to a spatially constant temporal
background gauge field representing Polyakov loop dynamics (the
Polyakov-linear-sigma model or the PLSM in short), the Lagrangian
reads \cite{Schaefer:2008ax}
\begin{eqnarray}\label{plsm}
\mathcal{L}=\mathcal{L}_{chiral}-\mathbf{\mathcal{U}}(\phi,\phi^*,T)
\end{eqnarray}
where we have separated the contribution of chiral degrees of
freedom and the Polyakov loop.

The chiral part of the Lagrangian, $\mathcal{L}_{chiral}=
\mathcal{L}_q+\mathcal{L}_m$\cite{Lenaghan:2000ey}\cite{Schaefer:2008hk},
of the $SU(3)_R \times SU(3)_L$ symmetric linear sigma model with
three quark flavors consists of the fermionic part
\begin{eqnarray}
\mathcal{L}_q=\sum_f \overline{\psi}_f(i\gamma^{\mu}
D_{\mu}-gT_a(\sigma_a+i \gamma_5 \pi_a))\psi_f \label{lfermion}
\end{eqnarray}
and the purely mesonic contribution
\begin{eqnarray}
\mathcal{L}_m &=&
\mathrm{Tr}(\partial_{\mu}\Phi^{\dag}\partial^{\mu}\Phi-m^2
\Phi^{\dag} \Phi)-\lambda_1 [\mathrm{Tr}(\Phi^{\dag} \Phi)]^2
\nonumber\\&& -\lambda_2 \mathrm{Tr}(\Phi^{\dag}
\Phi)^2+c[\mathrm{Det}(\Phi)+\mathrm{Det}(\Phi^{\dag})]
\nonumber\\&&+\mathrm{Tr}[H(\Phi+\Phi^{\dag})],\label{lmeson}
\end{eqnarray}
the sum is over the three flavors (f=1,2,3 for u, d, s). In the
above equation we have introduced a flavor-blind Yukawa coupling $g$
of the quarks to the mesons and the coupling of the quarks to a
background gauge field $A_{\mu}=\delta_{\mu 0}A_0$ via the covariant
derivative $D_{\mu}=\partial_{\mu}-i A_{\mu}$. The $\Phi$ is a
complex $3 \times 3$ matrix and is defined in terms of the scalar
$\sigma_a$ and pseudoscalar $\pi_a$ meson nonets,
\begin{eqnarray}
\Phi=T_a(\sigma_a+i\pi_a ).
\end{eqnarray}
The $3 \times 3$ matrix $H$ breaks the symmetry explicitly and is
chosen as
\begin{eqnarray}
H=T_a h_a,
\end{eqnarray}
where $h_a$ are nine external fields. The $T_a=\lambda_a/2$ are the
generators of the $U(3)$ symmetry, $\lambda_a$ are the Gell-Mann
matrices with $\lambda_0=\sqrt{\frac{2}{3}}\textbf{1}$. The $T_a$
are normalized to $\mathrm{Tr}(T_a T_b)=\delta_{ab}/2$ and obey the
$U(3)$ algebra with $[T_a,T_b]=i f_{abc}T_c$ and
$\{T_a,T_b\}=d_{abc} T_c$ respectively, here $f_{abc}$ and $d_{abc}$
for $a,b,c=1,...,8$ are the standard antisymmetric and symmetric
structure constants of $SU(3)$ group and
\begin{eqnarray}
f_{ab0}\equiv 0, \qquad d_{ab0}=\sqrt{\frac{2}{3}}\delta_{ab}.
\end{eqnarray}

In Eq.(\ref{lmeson}), $m^2$ is the tree-level mass square of the
fields in the absence of symmetry breaking, $\lambda_1$ and
$\lambda_2$ are two possible quartic coupling constants, and $c$ is
the cubic coupling constant. The terms in the first line of
Eq.(\ref{lmeson}) are invariant under $U(3)_R \times U(3)_L \cong
U(3)_V \times U(3)_A$ transformations, the determinant terms are
invariant under $SU(3)_R \times SU(3)_L\cong SU(3)_V \times
SU(3)_A$, but break the $U(1)_A$ symmetry explicitly, these terms
arise from the $U(1)_A$ anomaly of the QCD vacuum. The last term in
Eq.(\ref{lmeson}) breaks the axial and the $SU(3)_V$ vector
symmetries explicitly. A non-vanishing vacuum expectation value for
$\Phi$, $\langle \Phi \rangle \equiv T_a \bar{\sigma}_a$, breaks the
chiral symmetry spontaneously. Because the parity is not broken in
the vacuum, there are no non-vanishing vacuum expectation values for
fields $\pi_a$. The patterns of explicit symmetry breaking have been
described in detail in Ref.\cite{Lenaghan:2000ey} for the $U(3)_V
\times U(3)_A$ model, in this work we will constrain our study on
the case of explicit chiral symmetry breaking with $U(1)_A$ anomaly.
Since $\langle \Phi \rangle$ must carry the quantum numbers of the
vacuum, only the fields $\bar{\sigma}_a$ corresponding to the
diagonal generators of $U(3)$ can be nonzero. The same holds for
$h_a$, then the diagonal components $h_0$, $h_3$ and $h_8$ of the
explicit symmetry breaking term could be nonzero. Furthermore due to
the fact that the masses of the up and down quarks are approximately
equal, $m_{u}\simeq m_d$, and the strange quark mass $m_s$ is larger
than $m_{u}$ and $m_d$, in the following discussion, we restrict our
study to $h_0\neq 0$, $h_3=0$ and $h_8\neq 0 $. In this case the
$SU(3)_V \times SU(3)_A$ symmetry is explicitly broken to $SU(2)_V$
isospin symmetry, and the $U(1)_A$ subgroup of the $U(3)_A$ symmetry
is already explicitly broken by the instanton. Traditionally,
the linear sigma model does not have quark degrees of freedom, the nonzero quark masses
correspond to a term of the form Eq.(\ref{lmeson}) in the lagrangian,
where the matrix $H$ is proportional to the quark mass matrix,
the fields $H$ can be determined from the vacuum values for the
pion and kaon masses, as well as the pion and kaon decay constants,
the values for the constants of proportionality could be fixed at
bare quark masses for $u,d$ and $s$ quarks,
for example $m_q=10$ MeV, $m_s=150$ MeV\cite{Lenaghan:2000kr}.
This means that the parameters in the purely mesonic Lagrangian carries the information
of the bare quark mass, in order to avoid the double counting, such
model do not introduce bare quark masses in Lagrangian (\ref{lfermion}).

The quantity $\mathbf{\mathcal{U}}(\phi,\phi^*,T)$ is the
Polyakov-loop effective potential expressed by the dynamics of the
traced Polyakov loop
\begin{eqnarray}
\phi=(\mathrm{Tr}_c L)/N_c, \qquad \phi^*=(\mathrm{Tr}_c
L^{\dag})/N_c.
\end{eqnarray}
The Polyakov loop $L$ is a matrix in color space and explicitly
given by
\begin{eqnarray}
L(\vec{x})=\mathcal{P}\mathrm{exp}\left[i\int_0^{\beta}d \tau
A_4(\vec{x},\tau)\right],
\end{eqnarray}
with $\beta=1/T$ being the inverse of temperature and $A_4=iA^0$. In
the so-called Polyakov gauge, the Polyakov-loop matrix can be given
as a diagonal representation \cite{Fukushima:2003fw}. The coupling
between Polyakov loop and quarks is uniquely determined by the
covariant derivative $D_{\mu}$ in the PLSM Lagrangian in
Eq.(\ref{plsm}), and in the chiral limit, this Lagrangian is
invariant under the chiral flavor group, just like the original QCD
Lagrangian. The trace of the Polyakov-loop, $\phi$ and its conjugate
$\phi^*$ can be treated as classical field variables in this work.

The temperature dependent effective potential
$\mathbf{\mathcal{U}}(\phi,\phi^*,T)$ is used to reproduced the
thermodynamical behavior of the Polyakov loop for the pure gauge
case in accordance with lattice QCD data, and it has the $Z(3)$
center symmetry like the pure gauge QCD Lagrangian. In the absence
of quarks, we have $\phi=\phi^*$ and the Polyakov loop is taken as
an order parameter for deconfinement. For low temperatures,
$\mathbf{\mathcal{U}}$ has a single minimum at $\phi=0$, while at
high temperatures it develops a second one which turns into the
absolute minimum above a critical temperature $T_0$, and the $Z(3)$
center symmetry is spontaneously broken. In this paper, we will use
the potential $\mathbf{\mathcal{U}}(\phi,\phi^*,T)$ proposed in
Ref.\cite{Ratti:2005jh}, which has a polynomial expansion in $\phi$
and $\phi^*$:
\begin{eqnarray}
\frac{\mathbf{\mathcal{U}}(\phi,\phi^*,T)}{T^4}=-\frac{b_2(T)}{2}|\phi|^2-\frac{b_3
}{6}(\phi^3+\phi^{*3})+\frac{b_4}{4}(|\phi|^2)^2,
\end{eqnarray}
with
\begin{eqnarray}
b_2(T)=a_0+a_1\left(\frac{T_0}{T}\right)+a_2\left(\frac{T_0}{T}\right)^2+a_3\left(\frac{T_0}{T}\right)^3.
\end{eqnarray}
A precision fit of the constants $a_i,b_i$ is performed to reproduce
the lattice data for pure gauge theory thermodynamics and the
behavior of the Polyakov loop as a function of temperature. The
corresponding parameters are
\begin{eqnarray}
a_0=6.75,\qquad a_1=-1.95,\qquad a_2=2.625, \nonumber \\
 a_3=-7.44,\qquad
b_3=0.75,\qquad b_4=7.5.
\end{eqnarray}
The critical temperature $T_0$ for deconfinement in the pure gauge
sector is fixed at $270$ MeV, in agreement with the lattice results.

The chiral part of the model involves seven parameter $m^2$,
$\lambda_1$, $\lambda_2$, $c$, $g$, $h_0$, $h_8$ and two unknown
condensates $\bar{\sigma}_0$ and $\bar{\sigma}_8$, which we take
from \cite{Schaefer:2008hk}, where they were fixed to reproduce some
physical quantities in the hadronic sector. The vacuum condensates
$\bar{\sigma}_0$ and $\bar{\sigma}_8$ are members of the scalar
($J^p=0^+$) nonet and both contain the up-down quark condensate and
the strange quark condensate. It is more convenient to convert the
condensates into a pure up-down quark condensate and strange
condensate, this is achieved by an orthogonal basis transformation
from the original octet-singlet basis ($\bar{\sigma}_0$,
$\bar{\sigma}_8$) to the up-down quark condensate $\sigma_x$ and the
strange quark condensate $\sigma_y$ basis (for simplicity, we ignore
the bar over the up-down quark condensate $\sigma_x$ and the strange
quark condensate $\sigma_x$ ),
\begin{subequations}
\begin{eqnarray}
\sigma_x &=&
\sqrt{\frac{2}{3}}\bar{\sigma}_0 +\frac{1}{\sqrt{3}}\bar{\sigma}_8, \\
\sigma_y &=&
\frac{1}{\sqrt{3}}\bar{\sigma}_0-\sqrt{\frac{2}{3}}\bar{\sigma}_8.
\end{eqnarray}
\end{subequations}
As a result, when chiral symmetry breaks spontaneously, the fields
$\langle \Phi \rangle$ acquires a non-vanishing vacuum expectation
value, the constituent quark masses for $u$, $d$ and $s$ are defined
to be
\begin{eqnarray}\label{qmass}
m_q &=& g \sigma_x/2 \qquad for \qquad u,d, \\
m_s &=& g \sigma_y/\sqrt{2} \qquad for \qquad s,\label{sqmass}
\end{eqnarray}
and the light quark sector decouples from the strange quark sector.
At zero temperature and zero chemical potential, the parameters of
the Lagrangian are fixed in a way that these masses agree with the
observed value of pion mass $m_{\pi}=138$ MeV and the most commonly
accepted value for sigma mass $m_{\sigma}=600$ MeV. The values of
the condensates are determined from the pion and kaon decay
constants by means of the partially conserved axial-vector current
relation (PCAS),
\begin{eqnarray}
\sigma_x=f_{\pi}, \qquad \sigma_y=\frac{1}{\sqrt{2}}(2 f_K-f_{\pi}),
\end{eqnarray}
and the decay constants, $f_{\pi}=92.4$ MeV and $f_K=113$ MeV. The
Yukawa coupling is usually fixed by requirement that the constituent
quark mass in the vacuum, for $u$ and $d$ quarks, is about $1/3$ of
the nucleon mass, which gives $g\simeq 6.5$. The strange constituent
quark mass is predicted to be $m_s\simeq 433$ MeV. The six
parameters $m^2$, $\lambda_1$, $\lambda_2$, $c$, $h_x$ and $h_y$ are
set for $m_{\sigma}=600$ MeV with the axial $U(1)_A$ anomaly:
$m^2=(342.52 \mathrm{MeV})^2$ , $\lambda_1=1.40$, $\lambda_2=46.48$,
$c=4807.84$ MeV, $h_x=(120.73 \mathrm{ MeV})^3$ and $h_y=(336.41
\mathrm{ MeV})^3$\cite{Schaefer:2008hk}.

\section{Mean field approximation}

The standard approach for dealing with the thermodynamics of
variable particles is via the grand canonical ensemble. Let us
consider a spatially uniform system in thermodynamical equilibrium
at temperature $T$ and quark chemical potential $\mu_f (f=u,d,s)$.
In general, the grand partition function reads
\begin{eqnarray}
\mathcal{Z}&=& \mathrm{Tr exp}[-(\hat{\mathcal{H}}-\sum_{f=u,d,s}
\mu_f \hat{\mathcal{N}}_f)/T] \nonumber \\
&=& \int\prod_a \mathcal{D} \sigma_a \mathcal{D} \pi_a \int
\mathcal{D}\psi \mathcal{D} \bar{\psi} \mathrm{exp} \left[ \int_x
(\mathcal{L}+\sum_{f=u,d,s} \mu_f \bar{\psi}_f \gamma^0 \psi_f )
\right],
\end{eqnarray}
where $\int_x\equiv i \int^{1/T}_0 dt \int_V d^3x$ and $V$ is the
volume of the system. The $u$ and $d$ quark chemical potentials is
\begin{eqnarray}
\mu_u=\frac{\mu_B}{3}+\frac{\mu_I}{2}+\frac{1}{3}\mu_Y, \qquad
\mu_d=\frac{\mu_B}{3}-\frac{\mu_I}{2}+\frac{1}{3}\mu_Y, \qquad
\mu_s=\frac{\mu_B}{3}-\frac{2}{3}\mu_Y
\end{eqnarray}
where $\mu_B$, $\mu_I$ and $\mu_Y$ are the baryon, isospin and
hypercharge chemical potentials respectively. In general, the three
quark chemical potentials are independent, but in the following
discussion, we assume the $SU(2)_V$ isospin symmetry by neglecting
the slight mass difference between an $u-$ and $d-$ quark. Then the
light quark chemical potentials become equal, $\mu_q
\equiv\mu_u=\mu_d$, and there is no effect of the isospin density
since the isospin chemical potential $\mu_I$ is zero. Also for
simplicity and confronting our results with the other models, such
as LSM and PNJL models with three quark flavors, we will
only consider symmetric quark matter and define a uniform chemical
potential $\mu\equiv \mu_q=\mu_s$ for $\mu_Y=0$.

We evaluate the partition function in the mean-field approximation
similar to \cite{Schaefer:2008hk}\cite{Scavenius:2000qd}. Thus we
replace the meson fields by their expectation values in the action.
In other words, we neglect both quantum and thermal fluctuations of
the meson fields. The quarks and antiquarks are retained as quantum
fields. The integration over the fermions yields a determinant which
can be calculated by standard methods\cite{Kapusta:2006pm}. This
generates an effective potential for the mesons. Finally, we obtain
the thermodynamical potential density as
\begin{eqnarray}\label{potential}
\Omega(T,\mu_f)=\frac{-T \mathrm{ln}
\mathcal{Z}}{V}=U(\sigma_x,\sigma_y)+\mathbf{\mathcal{U}}(\phi,\phi^*,T)+\Omega_{\bar{\psi}
\psi},
\end{eqnarray}
with the quarks and antiquarks contribution
\begin{eqnarray}
\Omega_{\bar{\psi} \psi} &=& -2 T N_q \int \frac{d^3\vec{p}}{(2
\pi)^3} \{ \mathrm{ln} [ 1+3(\phi+\phi^* e^{-(E_q-\mu)/T})\times
e^{-(E_q-\mu)/T}+e^{-3 (E_q-\mu)/T}] \nonumber \\&&  +\mathrm{ln} [
1+3(\phi^*+\phi e^{-(E_q+\mu)/T})\times e^{-(E_q+\mu)/T}+e^{-3
(E_q+\mu)/T}] \} \nonumber \\&& -2 T N_s \int \frac{d^3\vec{p}}{(2
\pi)^3} \{ \mathrm{ln} [ 1+3(\phi+\phi^* e^{-(E_s-\mu)/T})\times
e^{-(E_s-\mu)/T}+e^{-3 (E_s-\mu)/T}] \nonumber \\&&  +\mathrm{ln} [
1+3(\phi^*+\phi e^{-(E_s+\mu)/T})\times e^{-(E_s+\mu)/T}+e^{-3
(E_s+\mu)/T}] \}.
\end{eqnarray}
Here, $N_q=2$, $N_s=1$, and $E_q=\sqrt{\vec{p}^2+m_q^2}$ is the
valence quark and antiquark energy for $u$ and $d$ quarks, for
strange quark $s$, it is $E_s=\sqrt{\vec{p}^2+m_s^2}$, and $m_q$,
$m_s$ is the constituent quark mass for $u$, $d$ and $s$ in
Eqs.(\ref{qmass})(\ref{sqmass}). The purely mesonic potential is
\begin{eqnarray}
U(\sigma_x,\sigma_y)=\frac{m^2}{2} (\sigma^2_x+\sigma^2_y)-h_x
\sigma_x-h_y \sigma_y-\frac{c}{2\sqrt{2}} \sigma^2_x \sigma_y
+\frac{\lambda_1}{2} \sigma^2_x \sigma^2_y +\frac{1}{8} (2 \lambda_1
+\lambda_2)\sigma^4_x + \frac{1}{4} (\lambda_1+\lambda_2)\sigma^4_y.
\end{eqnarray}
Minimizing the thermodynamical potential in Eq.(\ref{potential})
with respective to $\sigma_x$, $\sigma_y$, $\phi$ and $\phi^*$, we
obtain a set of equations of motion
\begin{eqnarray}
\frac{\partial \Omega}{\partial \sigma_x}=0, \qquad \frac{\partial
\Omega}{\partial \sigma_y}=0, \qquad \frac{\partial \Omega}{\partial
\phi}=0, \qquad \frac{\partial \Omega}{\partial \phi^*}=0.
\end{eqnarray}
The set of equations can be solved for the fields as functions of
temperature $T$ and chemical potential $\mu$, and the solutions of
these coupled equations determine the behavior of the chiral order
parameter $\sigma_x$, $\sigma_y$ and the Polyakov loop expectation
values $\phi$, $\phi^*$ as a function of $T$ and $\mu$.

\section{$T-\mu$ Phase diagram}

We now explore the phase diagram of the Polyakov linear sigma model
at finite temperature and density. There are three different
critical temperatures, i.e, $T_c^{q}$ for the chiral phase
transition of $u,d$ quarks, $T_c^{s}$ for the chiral phase
transition of $s$ quark, and $T_c^d$ for the deconfinement phase
transition. In order to locate the critical temperatures, we define
the temperature derivatives of the condensates: $\sigma_x'=\partial
\sigma_x /
\partial T $, $\sigma_y'=\partial \sigma_y / \partial T $,
$\phi'=\partial \phi /
\partial T$ and ${\phi ^{*}}'=\partial {\phi ^{*}} / \partial T$.
In the following, we will show the double peak structure of the
temperature derivatives of the condensates and explain explicitly
how we determine the critical temperatures.

\subsection{Phase transitions at zero density}

\begin{figure}[thbp]
\epsfxsize=7.5 cm \epsfysize=6.5cm
\epsfbox{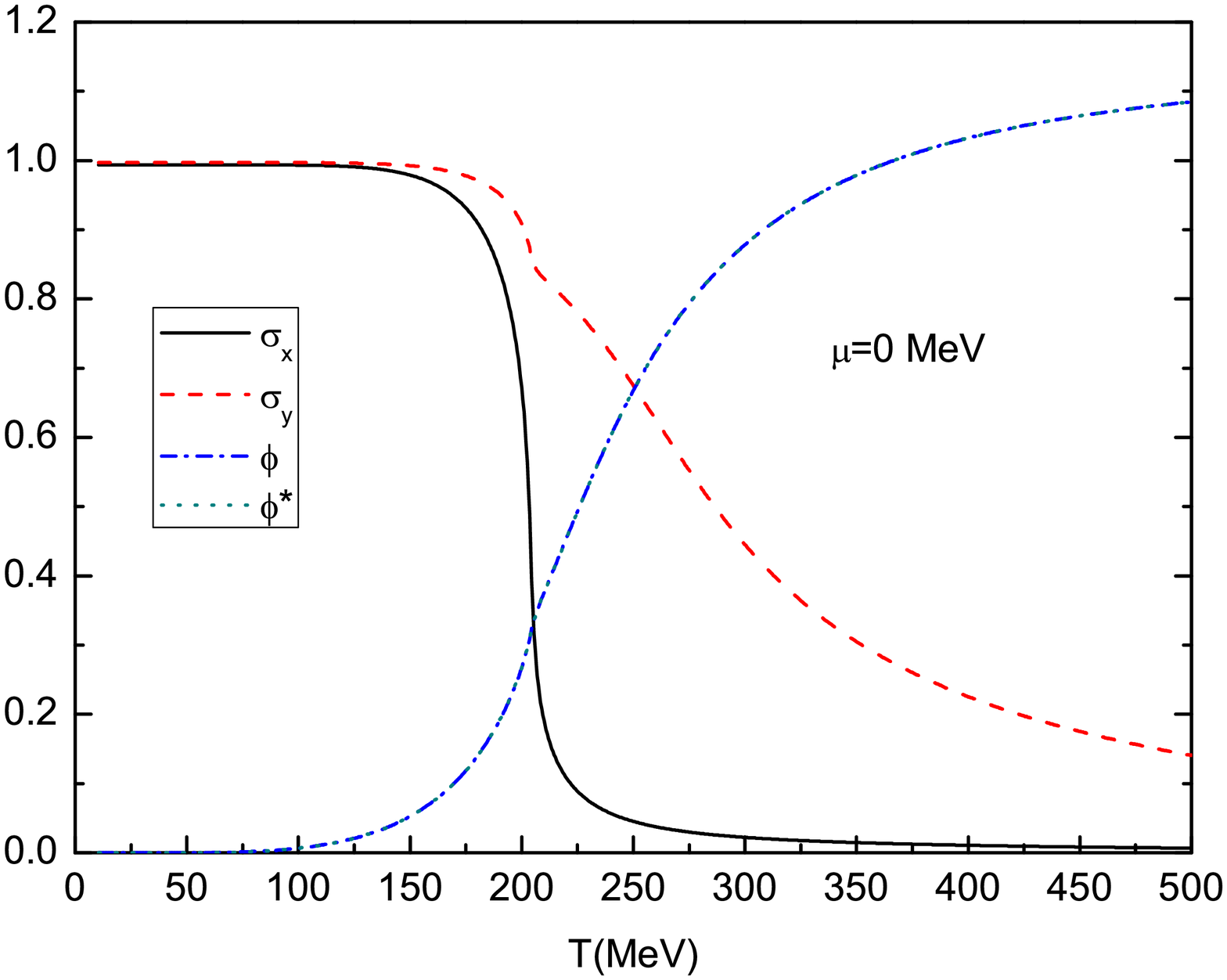}\hspace*{0.1cm} \epsfxsize=7.5 cm
\epsfysize=6.5cm \epsfbox{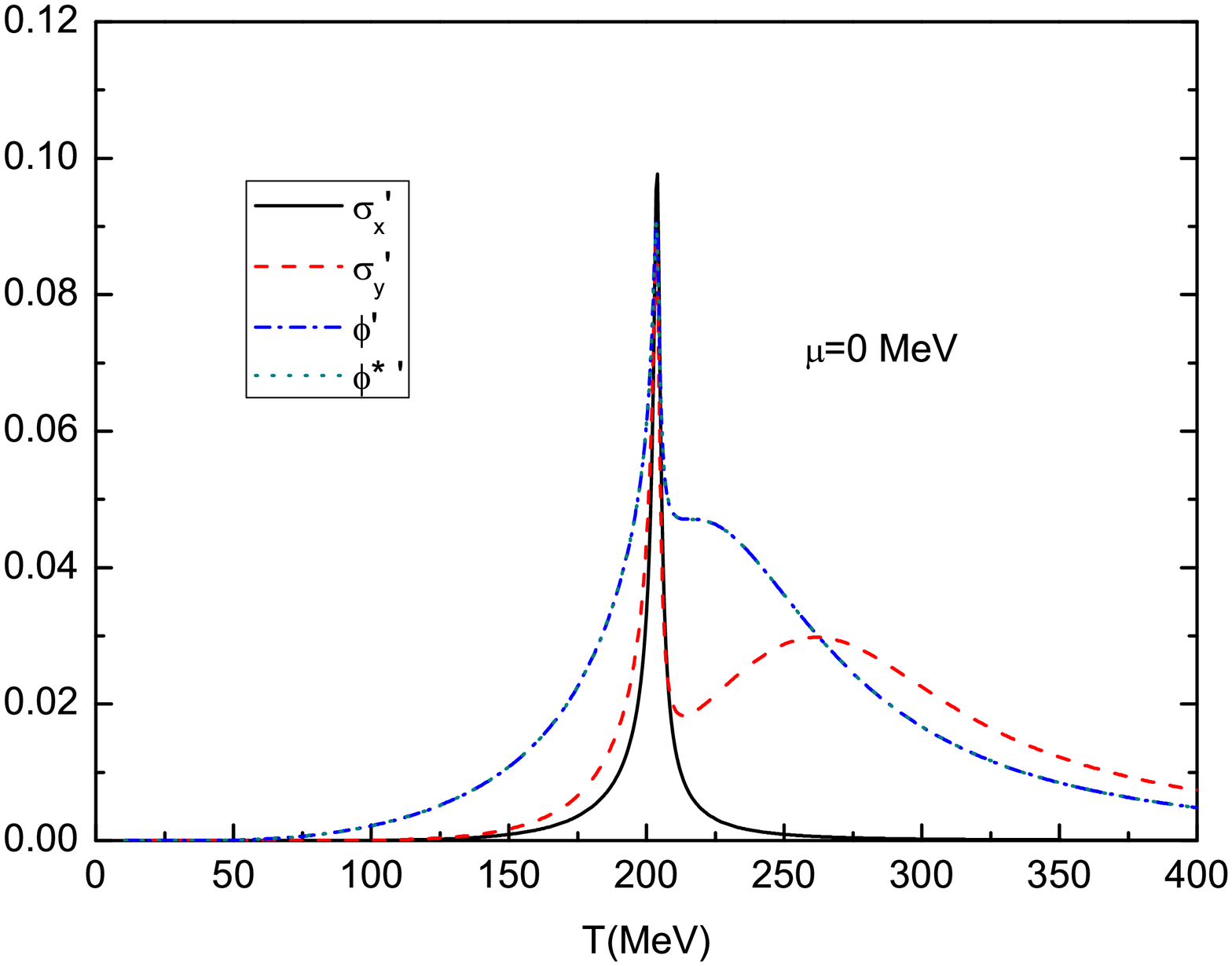}
\vskip -0.05cm \hskip 0.15 cm \textbf{( a ) } \hskip 6.5 cm \textbf{( b )} \\
 \caption{(a) The normalized chiral condensate $\sigma_x$, $\sigma_y$ and
the Polyakov loop $\phi$, $\phi^*$ as a function of temperature for
$\mu = 0$ MeV. (b) Temperature derivatives of the normalized chiral
condensate $\sigma_x$, $\sigma_y$ and the Polyakov loop $\phi$,
$\phi^*$ as a function of temperature at $\mu = 0$ MeV. The Polyakov
variable and $\sigma_y$ are scaled by a factor of 6.}
\label{Fig:Fig01}
\end{figure}

We firstly investigate the phase transitions at zero density. In
Fig.\ref{Fig:Fig01}(a), the temperature dependence of the chiral
condensates $\sigma_x$ for $u$, $d$ quarks, the chiral condensate
$\sigma_y$ for $s$ quark and the Polyakov loop expectation value
$\phi$, $\phi^*$ at $\mu =0$ MeV is shown in relative units. Here
and in the following discussion, the chiral condensates are
normalized by their zero-temperature value: $\sigma_{x0}=92.4$ MeV
and $\sigma_{y0}=94.5$ MeV for $u$, $d$ and $s$ quarks,
respectively. The temperature derivatives of the condensates
$\sigma_x', \sigma_y', \phi'$ and ${\phi ^{*}}'$ are shown in
Fig.\ref{Fig:Fig01}(b).

The temperature behavior of the chiral condensates and Polyakov loop
condensate shows that the system experiences a crossover at zero
chemical potential. The temperature derivative of the chiral
condensate $\sigma_x$ for $u$ and $d$ quarks has only one peak at
$T\simeq 200 $ MeV. The temperature derivative of the chiral
condensates $\sigma_y$ for $s$ quark has two peaks, the left peak
coincides with the peak of $\sigma_x'$ at $T\simeq 200 $ MeV, the
right peak shows up at $T\simeq 260$ MeV. For vanishing chemical
potential, we find $\phi=\phi^*$, and $\phi\simeq 1.1$ for
$T\rightarrow \infty$. The temperature derivatives $\phi'$ and
${\phi^{*}}'$ shows one peak and one pseudo-peak, and the peak
coincides with the peak of $\sigma_x'$ at $T\simeq 200 $ MeV.

Now we show how we determine the critical temperature of the
crossover. When there is only one peak in the temperature derivative
of the condensate, the location of the peak gives the critical
temperature $T_c$. When there are two peaks in the temperature
derivative of the condensate, for chiral phase transition, the
critical temperature is determined by the peak temperature
corresponding to $\sigma_{x,y}(T)/\sigma_{x,y}(T=0)<1/2$, and for
deconfinement phase transition, the critical temperature is given by
the peak or pseudo peak temperature corresponding to
$\phi(T)/\phi(T\rightarrow \infty)>1/2$. This definition of the
critical temperature is different from that defined in
Ref.\cite{Fukushima:2008wg}.

The temperature derivatives of the condensates in
Fig.\ref{Fig:Fig01} (b) show that the deconfinement phase transition
happens at a higher critical temperature $T_c^d\simeq 220$ MeV, and
the chiral restoration occurs at a smaller critical temperature
$T_c^{q}\simeq 203.5$ MeV. This result is similar to that obtained
in the two-flavor Polyakov linear sigma model
Ref.\cite{Kahara:2008yg}, and different from that in the
three-flavor Nambu-Jona-Lasinio model with the Polyakov loop
\cite{Fukushima:2008wg}, where they find the simultaneous crossovers
around $T_c\simeq 200$ MeV. The critical temperature for the chiral
restoration of the strange quark is at $T_c^s\simeq 260$ MeV.

\subsection{Phase transitions at finite density}

\begin{figure}[thbp]
\epsfxsize=7.5 cm \epsfysize=6.5cm
\epsfbox{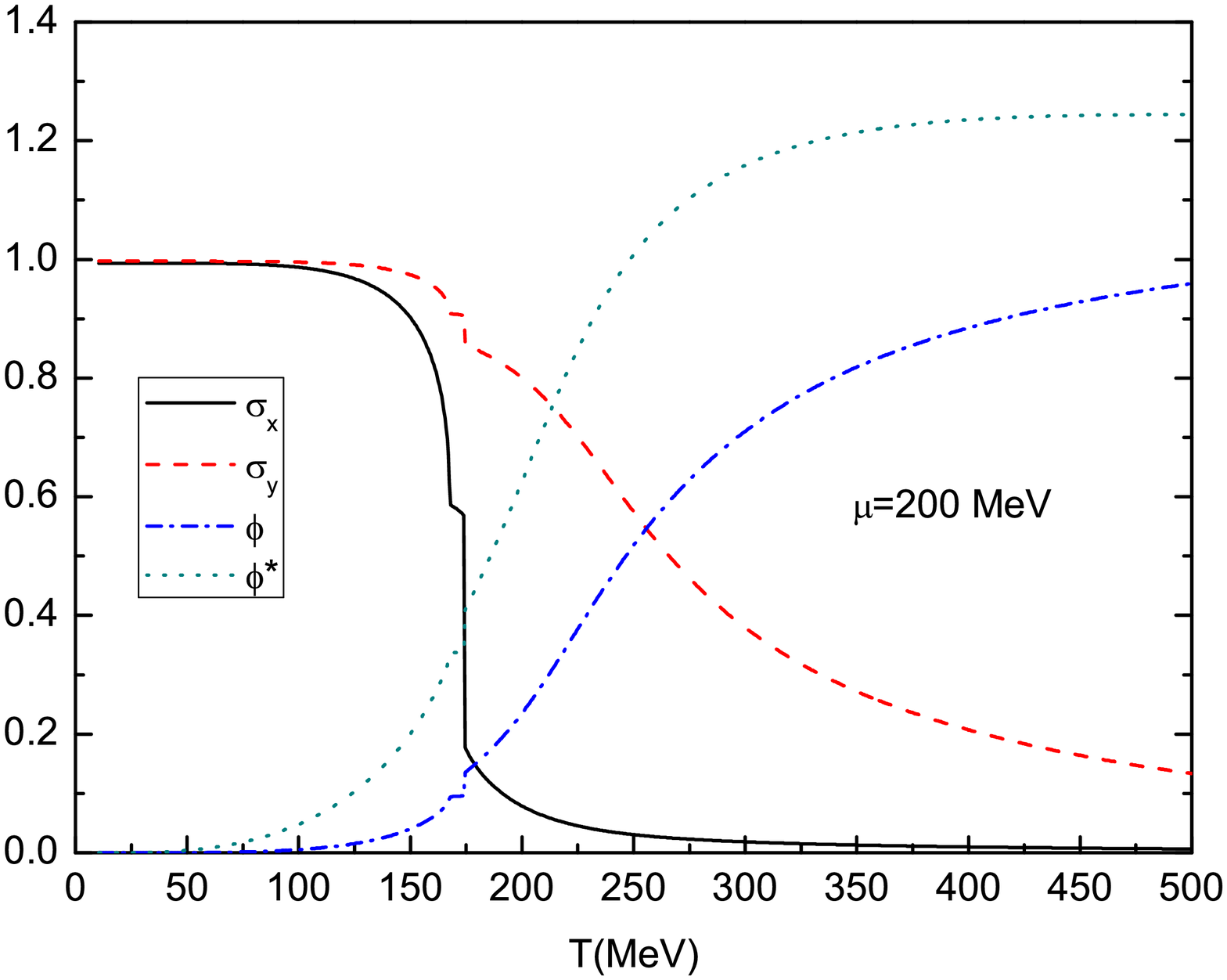}\hspace*{0.1cm} \epsfxsize=7.5 cm
\epsfysize=6.5cm \epsfbox{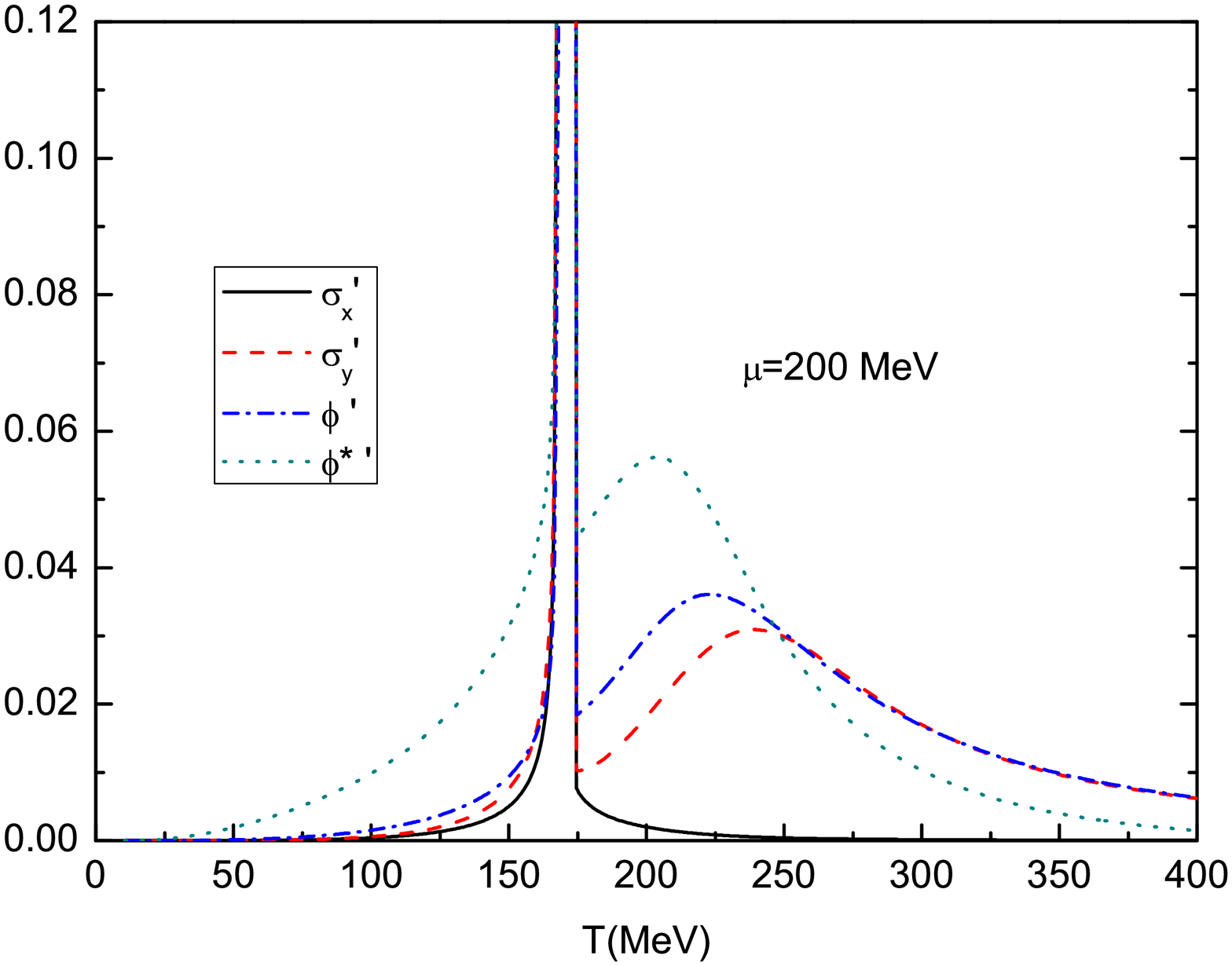}
\vskip -0.05cm \hskip 0.15 cm \textbf{( a ) } \hskip 6.5 cm \textbf{( b )} \\
 \caption{(a) The normalized chiral condensate $\sigma_x$, $\sigma_y$ and
the Polyakov loop $\phi$, $\phi^*$ as a function of temperature for
$\mu = 200$ MeV.  (b) Temperature derivatives of the normalized
chiral condensate $\sigma_x$, $\sigma_y$ and the Polyakov loop
$\phi$, $\phi^*$ as a function of temperature at $\mu = 200$ MeV.
The Polyakov variable and $\sigma_y$ are scaled by a factor of 6.}
\label{Fig:Fig02}
\end{figure}

For fixing the chemical potential at $200$ MeV, the order parameters
and their temperature derivatives as functions of temperature are
shown in Fig.\ref{Fig:Fig02} (a) and (b), respectively.

It is found that the temperature derivative of the chiral condensate
$\sigma_x$ for $u$ and $d$ quarks still has only one peak, at the
peak temperature $T_c^{q}\simeq 173$ MeV of $\sigma_x'$, the chiral
condensate jumps from $0.8$ to $0.15$, which indicates that the
chiral phase transition for the $u$ and $d$ quarks is of first order
phase transition.

The temperature derivative of chiral condensates $\sigma_y$ for $s$
quark has two peaks, the left peak coincides with the peak of
$\sigma_x'$ at $T_c^{q}\simeq 173 $ MeV, the right peak shows up at
$T_c^{s}\simeq 240$ MeV.

The values of the Polyakov loop $\phi$ and $\phi^*$ are different
for nonvanishing chemical potential, due to the fact that the free
energies of quark and antiquarks are different in the finite
chemical potential \cite{Fukushima:2006uv,Abuki:2009dt}. When $\mu$
increases, the left peak moves down to a smaller temperature, which
is always coincident with $T_c^{q}$, and the right pseudo-peak at
zero chemical potential develops into a real peak and stays almost
at the same temperature $T_c^d\simeq 220$ MeV as that for $\mu=0$.

Here are several remarks on the phase transitions. The left peak of
$\sigma_y'$ and $\phi'({\phi^*}')$ always coincides with the peak of
$\sigma_x'$, which reflects the effective coupling among $\sigma_x$,
$\sigma_y$ and $\phi({\phi^*})$. However, the coupling among the
condensates does not necessarily induce the simultaneous phase
transitions. From the above analysis, we can see that in general
cases, the three phase transitions, i.e, the chiral restoration of
$u,d$ quarks, the chiral restoration of $s$ quark, and the
deconfinement phase transition are independent in the Polyakov
linear sigma model. However the lattice result shows the
simultaneous phase transitions of chiral restoration and
deconfinement, it should be very interesting to investigate how to
realize the simultaneous phase transitions in the Polyakov linear
sigma model, and it deserves further efforts on understanding the
correlation between the chiral phase transitions and deconfinement
phase transition.

\subsection{$T-\mu$ phase structure and the location of the critical end point}

\begin{figure}[thbp]
\epsfxsize=7.5 cm \epsfysize=6.5cm
\epsfbox{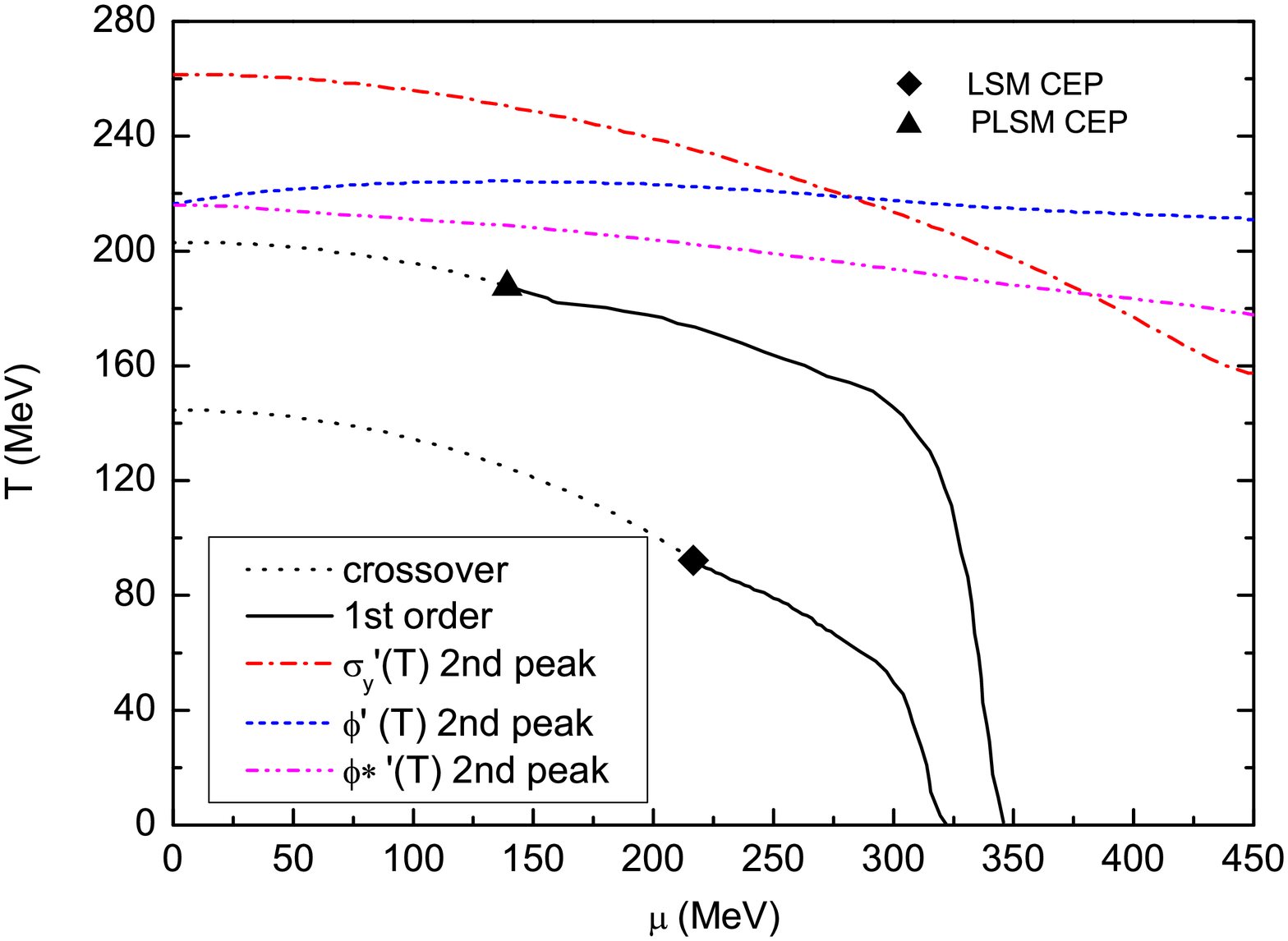}\hspace*{0.1cm} \epsfxsize=7.5 cm
\epsfysize=6.5cm \epsfbox{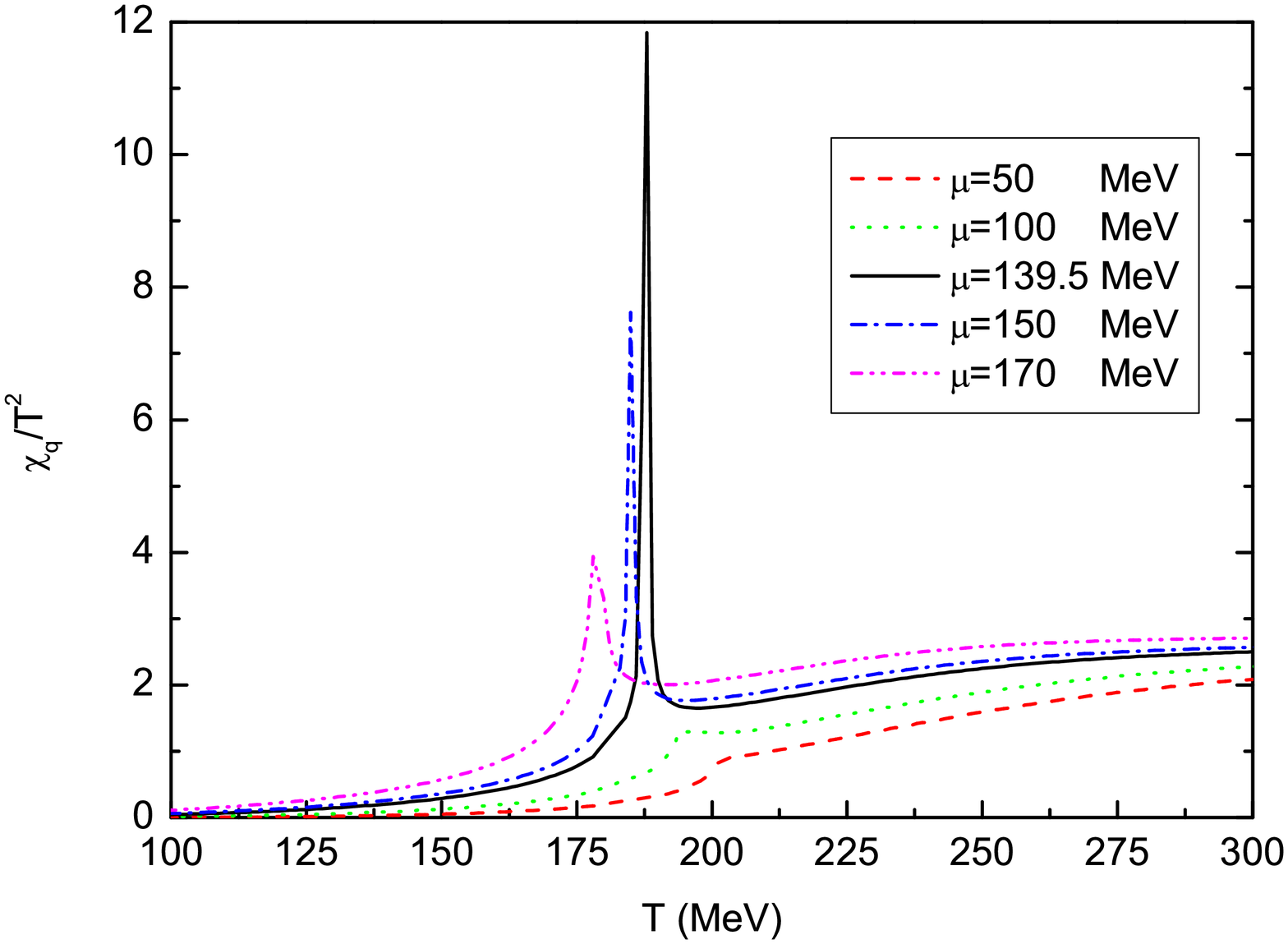}
\vskip -0.05cm \hskip 0.15 cm \textbf{( a ) } \hskip 6.5 cm \textbf{( b )} \\
 \caption{(a) The $T-\mu$ phase diagram in the Polyakov linear sigma model.
(b) the quark number susceptibility $\chi_q$ for different chemical
potentials as function of $T$. } \label{Fig:Fig03}
\end{figure}

We evaluate the chiral phase transitions of $u,d$ and $s$ quarks and
deconfinement phase transition at finite temperature and finite
density, and show the $T-\mu$ phase structure of the Polyakov linear
sigma model in Fig.\ref{Fig:Fig03} (a).

From Fig.\ref{Fig:Fig01} and Fig.\ref{Fig:Fig02}, we see that the
chiral phase transition for $s$ quark and the deconfinement phase
transition for the Polyakov loop are always crossover, for two light
flavors, there is a crossover in the low density region and a
first-order phase transition in the high-density region, and in the
middle exists a critical end point (CEP). In order to locate the
critical end point, we define the quark number susceptibility
$\chi_q=\partial^2 \Omega/\partial\mu^2$, which should be divergent
at the critical end point. In Fig. \ref{Fig:Fig03} (b), we plot the
quark number susceptibility as a function of the temperature for
different chemical potentials. For the Polyakov linear sigma model,
the result shows that the critical end point is around
$(T_E,\mu_E)=(188 ~{\rm MeV}, 139.5~{\rm MeV})$, which is close to
the lattice result $(T_E, \mu_E)=(162\pm 2~{\rm MeV}, \mu_E=120 \pm
13~{\rm MeV})$ \cite{Fodor:2004nz}. For the linear sigma model
without the Polyakov loop, the critical end point is located at
$(T_E, \mu_E) \simeq (92.5 {\rm MeV}, 216~{\rm MeV})$. The critical
chemical potential $\mu_E$ in PLSM is much lower than that in the
PNJL model with three quark flavors where the predicted critical end
point is $\mu_E > 300$ MeV\cite{Fu:2007xc,Ciminale:2007sr}.

The chiral phase transition for the strange quark and the
deconfinement phase transition in the $T-\mu$ plane are shown in
Fig.\ref{Fig:Fig03} (a) by the dash-dotted line and dotted line,
respectively. It is found that with the increase of chemical
potential, the critical temperature for strange quark to restore
chiral symmetry decreases. However, for the deconfinement phase
transition, with the increase of chemical potential, the
deconfinement critical temperature keeps almost a constant around
$220$ MeV. It can be seen that in the Polyakov linear sigma model,
there exists two-flavor quarkyonic phase \cite{McLerran:2007qj} at
low density, where the $u,d$ quarks restore chiral symmetry but
still in confinement, and three-flavor quarkyonic phase at high
density, where the $u,d,s$ quarks restore chiral symmetry but still
in confinement.

\section{Thermodynamic properties of the Polyakov linear sigma model}

In order to investigate the influence of the Polyakov loop on the
equilibrium thermodynamics, we calculate several thermodynamic
quantities. All information of the system is contained in the grand
canonical potential which is given by $\Omega$ in
Eq.(\ref{potential}), evaluated at the mean field level. The entropy
density is determined by taking the derivative of effective
potential with respect to the temperature, i.e,
\begin{eqnarray} \label{entropy}
s=-\partial \Omega(\phi)/\partial T .
\end{eqnarray}
As the standard treatment in lattice calculation, we introduce the
normalized pressure density $p$ which is normalized to vanish at
$T=\mu=0$ and the energy density $\varepsilon$ as
\begin{eqnarray}
p=-\Omega, \,\, \varepsilon=-p+ T s.
\end{eqnarray}
The equation of state $p(\varepsilon)$ is an important input into
hydrodynamics. The square of the speed of sound $C_s^2$ is related
to $p/\varepsilon$ and has the form of
\begin{equation}
C_s^2=\frac{{\rm d}p}{{\rm d}\varepsilon}=\frac{s}{T {\rm d}s/{\rm
d}T}=\frac{s}{C_v},
\end{equation}
where
\begin{eqnarray}
C_v=\partial \varepsilon/\partial T,
\end{eqnarray}
is the specific heat. At the critical temperature, the entropy
density as well as the energy density change most quickly with
temperature, thus one expect that $C_s^2$ should have a minimum at
$T_c$. The trace anomaly of the energy-momentum tensor ${\cal
T}^{\mu\nu}$
\begin{equation}\label{energymt}
\Delta=\frac{{\cal T}^{\mu\mu}}{T^4}\equiv \frac{\varepsilon-3
p}{T^4} =T\frac{\partial}{\partial T}(p/T^4)
\end{equation}
is a dimensionless quantity, which is also called the "interaction
measure".

\begin{figure}[thbp]
\epsfxsize=7.5 cm \epsfysize=6.5cm
\epsfbox{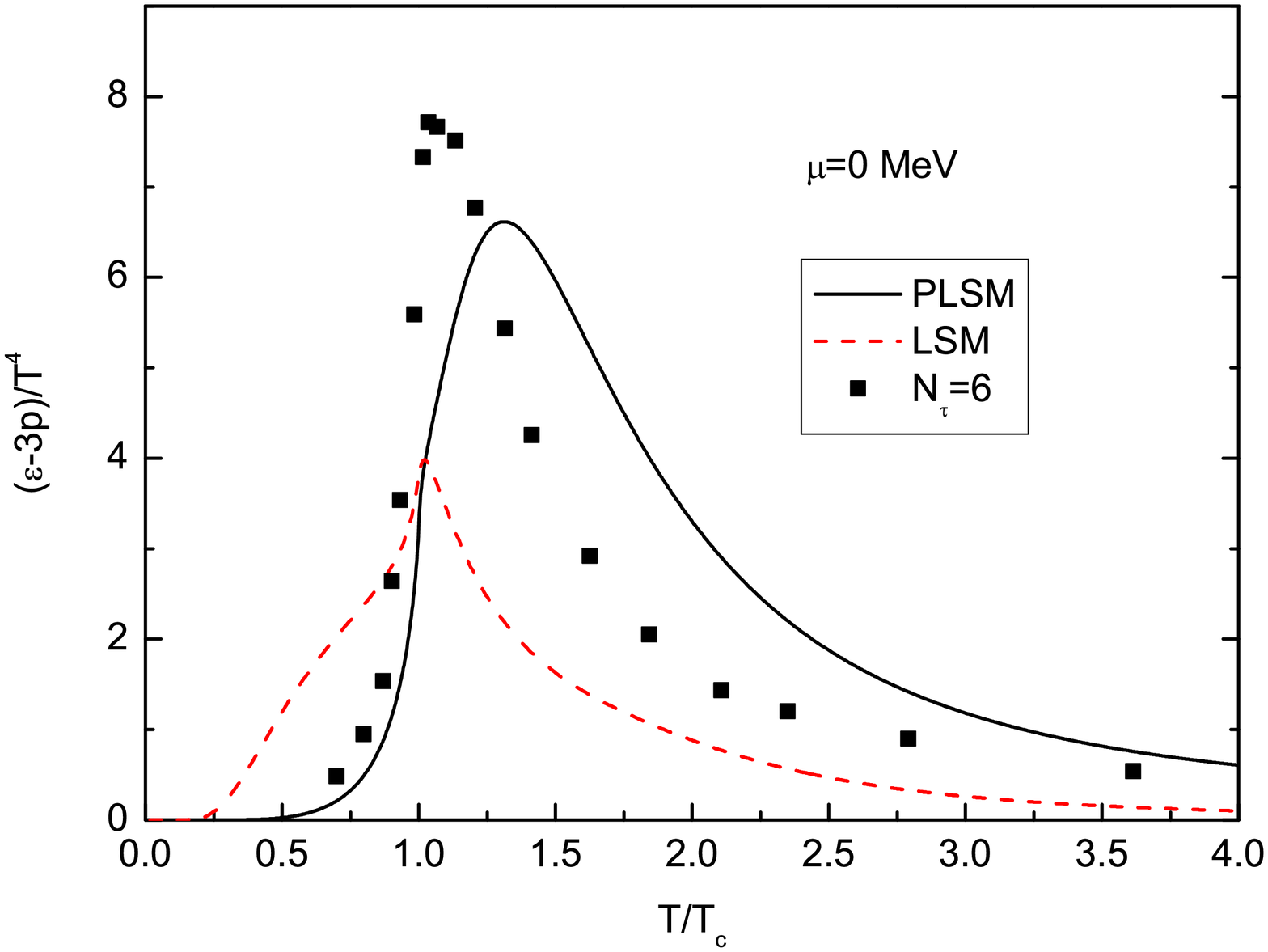}\hspace*{0.1cm} \epsfxsize=7.5 cm
\epsfysize=6.5cm \epsfbox{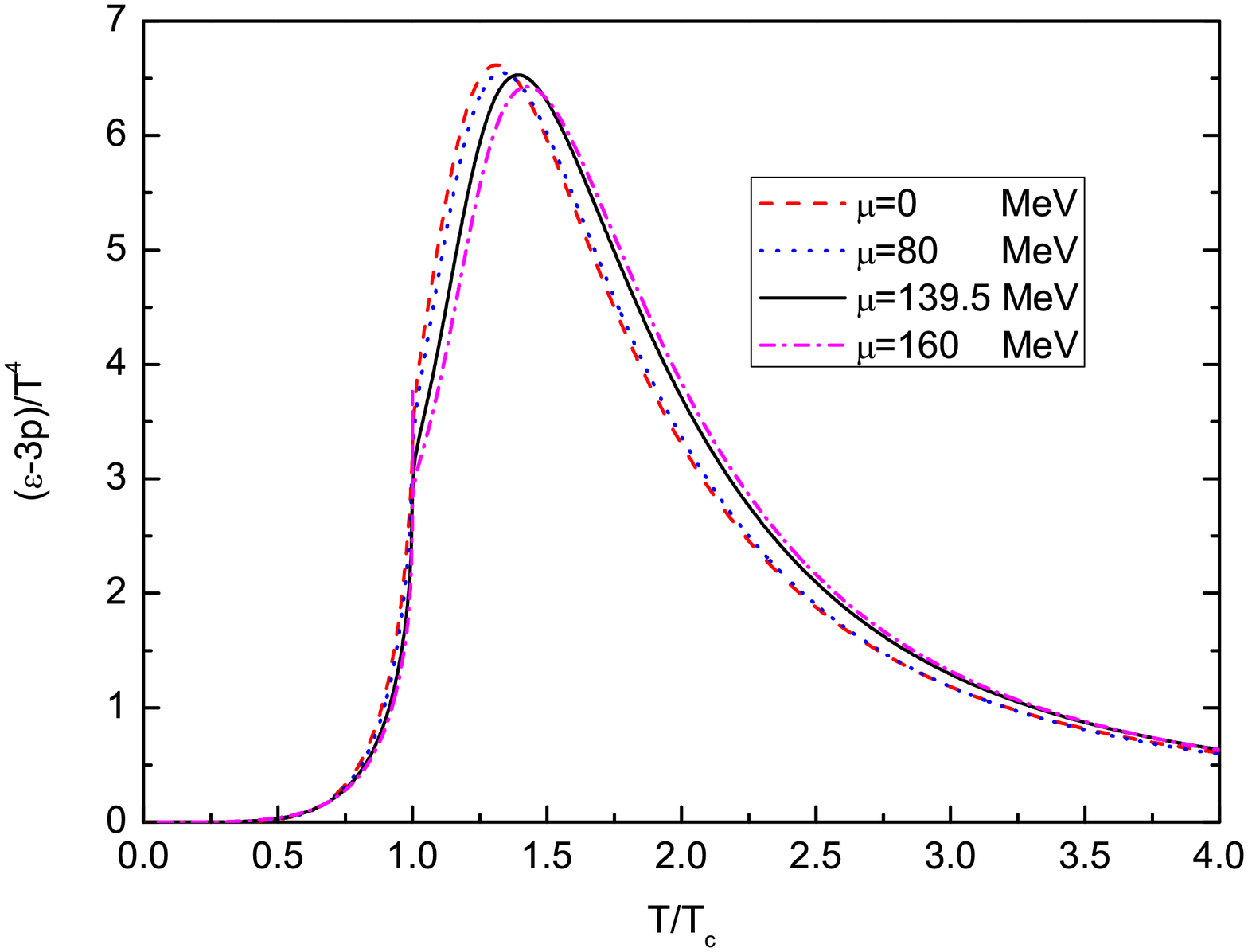}
\vskip -0.05cm \hskip 0.15 cm \textbf{( a ) } \hskip 6.5 cm \textbf{( b )} \\
 \caption{(a) The scaled trace anomaly $(\varepsilon-3p)/T^4$ for $\mu=0$
MeV. The Polyakov linear sigma model prediction (solid line) and the
linear sigma model prediction (dashed line) are compared with
$N_f=2+1$ lattice QCD data for $N_{\tau}=6$. Lattice data taken from
Ref.\cite{Cheng:2007jq}. (b) The scaled trace anomaly
$(\varepsilon-3p)/T^4$ in the Polyakov linear sigma model for
different chemical potentials as functions of $T/T_c$.}
\label{Fig:Fig04}
\end{figure}

\begin{figure}[thbp]
\epsfxsize=7.5 cm \epsfysize=6.5cm
\epsfbox{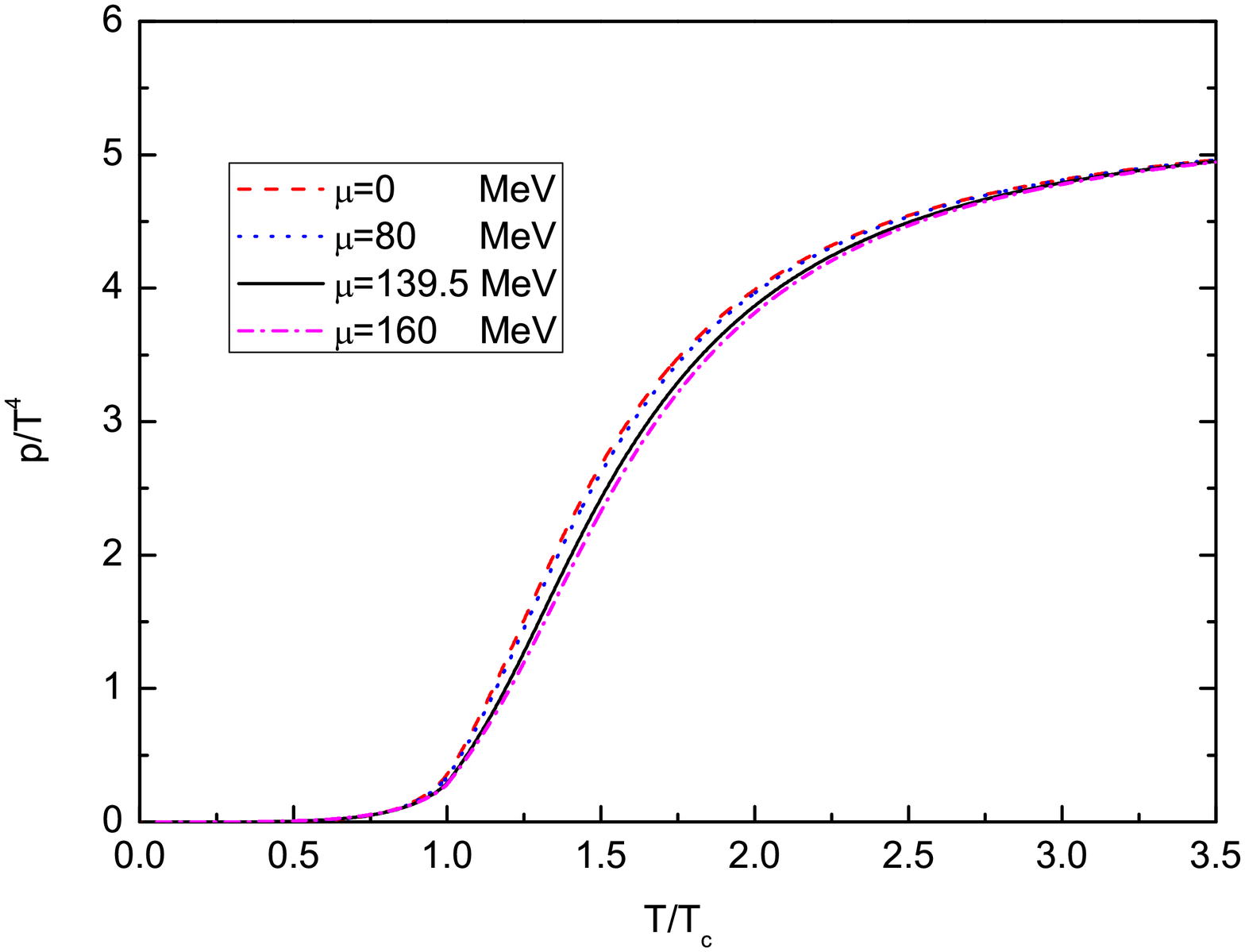}\hspace*{0.1cm} \epsfxsize=7.5 cm \epsfysize=6.5 cm
\epsfbox{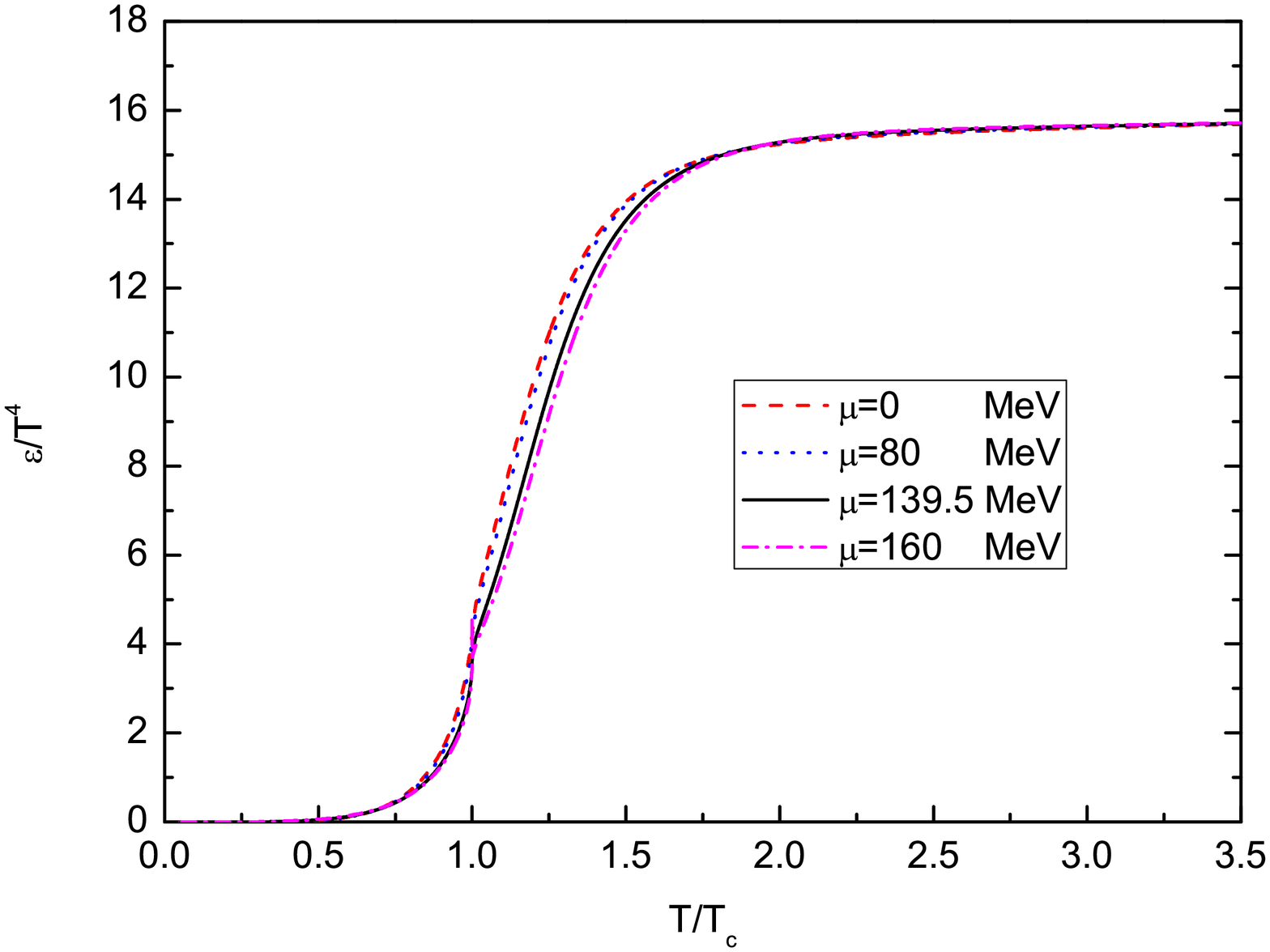}
\vskip -0.05cm \hskip 0.15 cm \textbf{( a ) } \hskip 6.5 cm \textbf{( b )} \\
 \caption{(a) The scaled pressure density $p/T^4$ in the Polyakov linear sigma
model for different chemical potentials as functions of $T/T_c$. (b)
The scaled energy density $\varepsilon/T^4$ in the Polyakov linear
sigma model for different chemical potentials as functions of
$T/T_c$. } \label{Fig:Fig05}
\end{figure}

\begin{figure}[thbp]
\epsfxsize=7.5 cm \epsfysize=6.5cm
\epsfbox{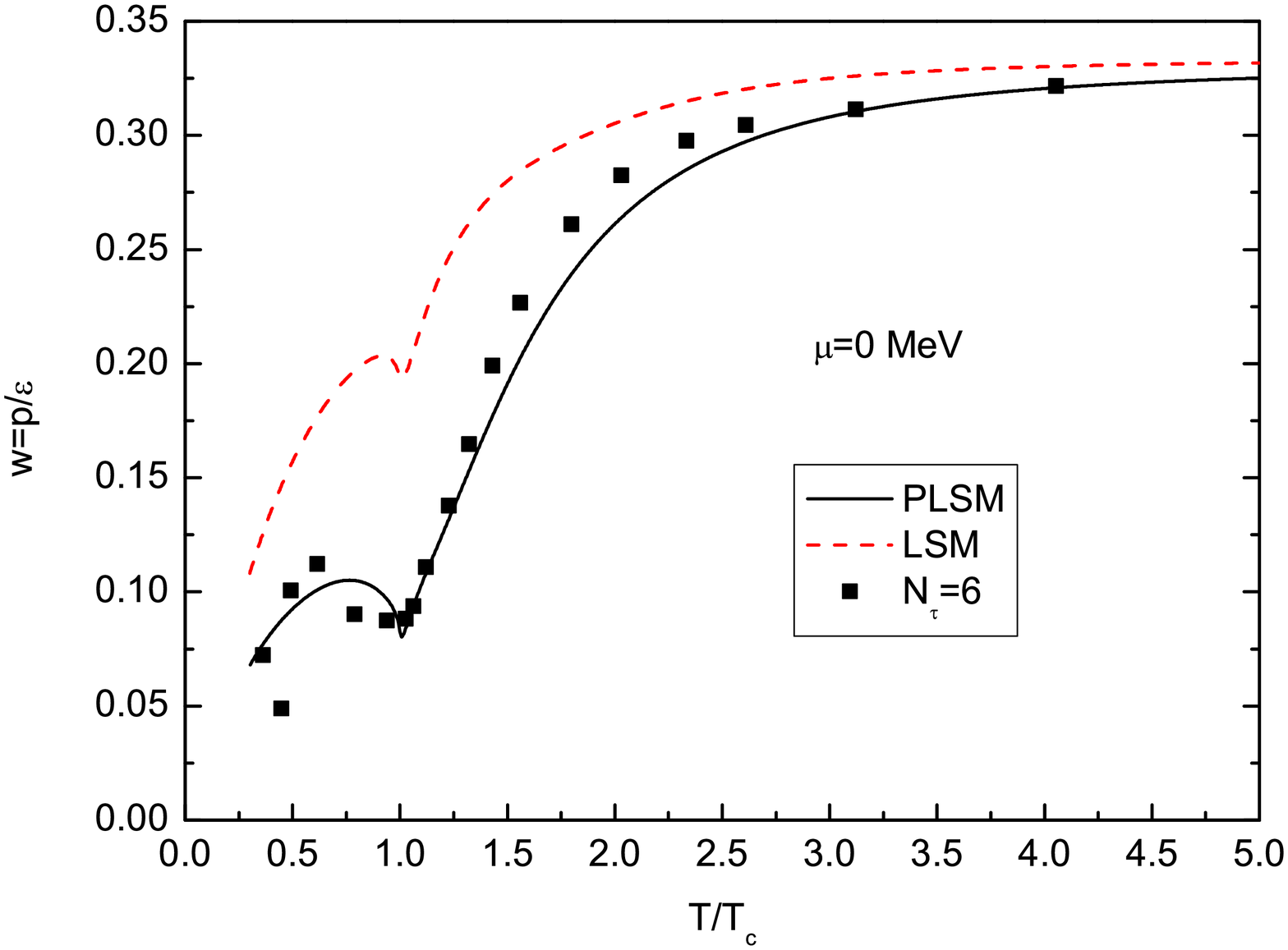}\hspace*{0.1cm} \epsfxsize=7.5 cm
\epsfysize=6.5cm \epsfbox{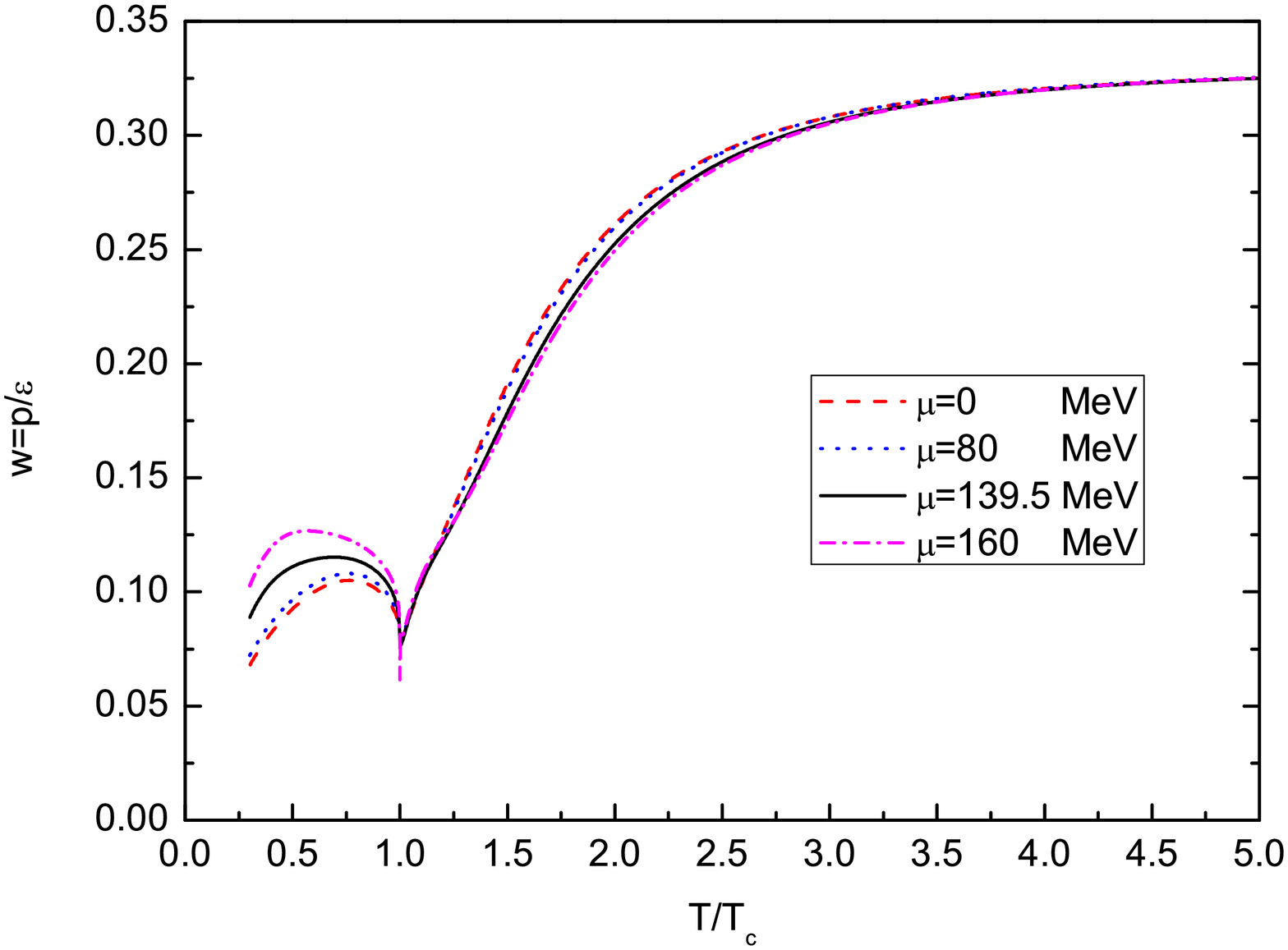}
\vskip -0.05cm \hskip 0.15 cm \textbf{( a ) } \hskip 6.5 cm \textbf{( b )} \\
 \caption{(a) The equation-of-state parameter $w(T)=p(T)/\varepsilon(T)$
for $\mu=0$ MeV. The Polyakov linear sigma model prediction (solid
line) and the linear sigma model prediction (dash line) are compared
with $N_f=2+1$ lattice QCD data for $N_{\tau}=6$. Lattice data taken
from Ref.\cite{Cheng:2007jq}. (b) The equation-of-state parameter
$w(T)=p(T)/\varepsilon(T)$ in the PLSM for different chemical
potentials as functions of $T/T_c$.} \label{Fig:Fig06}
\end{figure}

\begin{figure}[thbp]
\epsfxsize=7.5 cm \epsfysize=6.5cm
\epsfbox{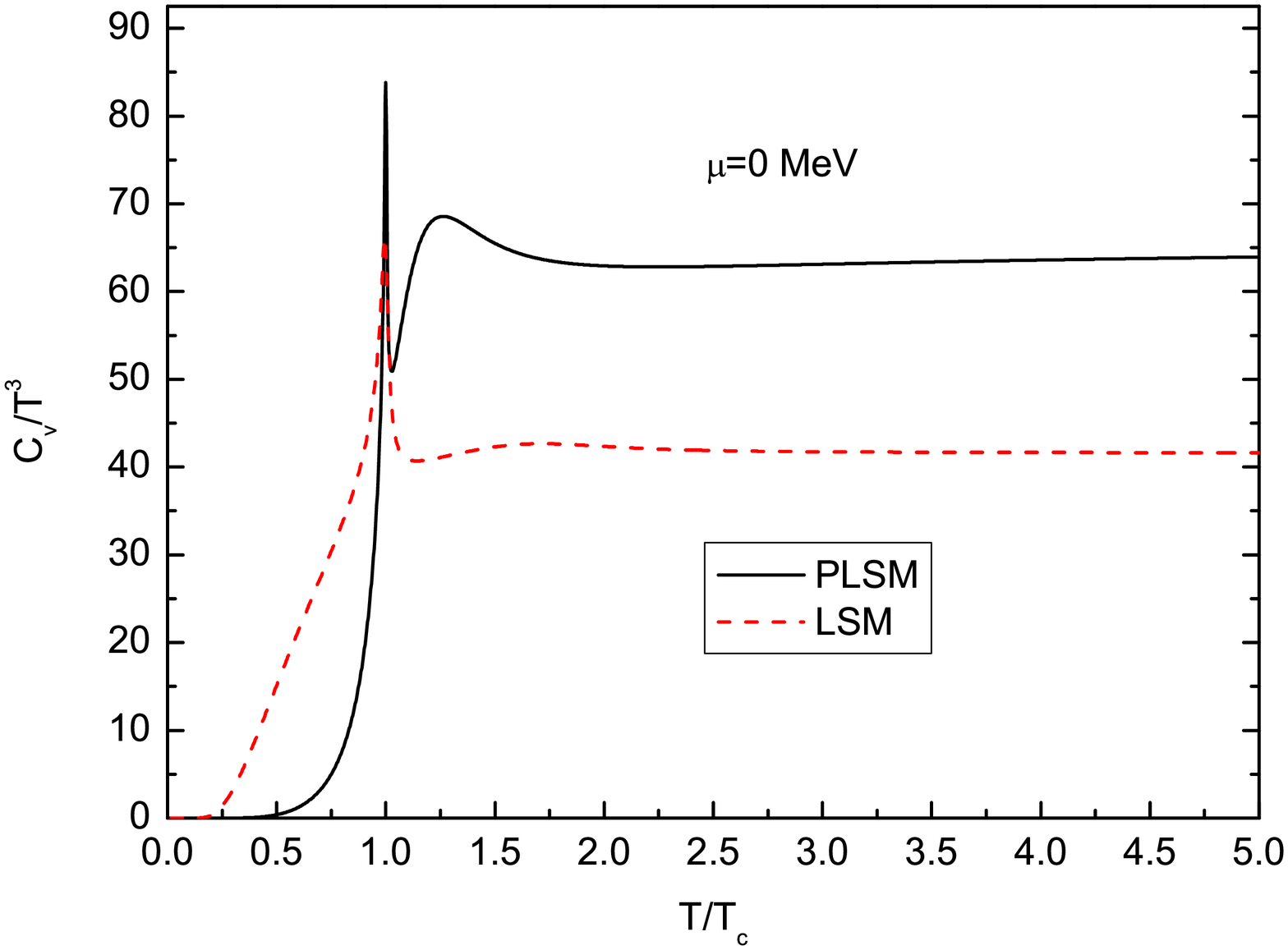}\hspace*{0.1cm} \epsfxsize=7.5 cm
\epsfysize=6.5cm \epsfbox{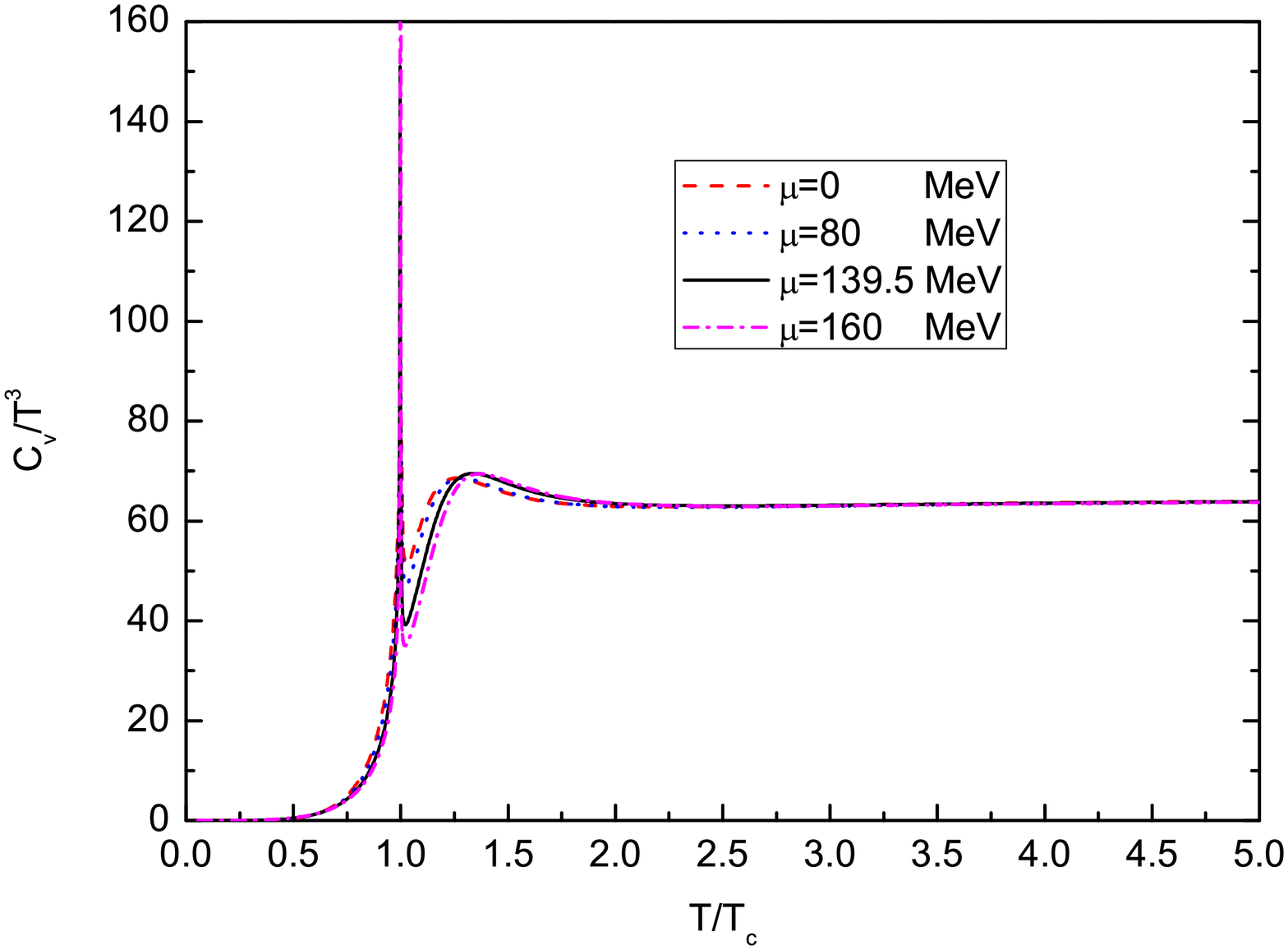}
\vskip -0.05cm \hskip 0.15 cm \textbf{( a ) } \hskip 6.5 cm \textbf{( b )} \\
 \caption{
(a) The specific heat $C_v$ as a function of the temperature for
$\mu=0$ MeV. The solid line denotes the Polyakov linear sigma model
prediction and the dashed line denotes the linear sigma model
prediction. (b)The specific heat $C_v$ in the PLSM for different
chemical potentials as functions of $T/T_c$.} \label{Fig:Fig07}
\end{figure}

\begin{figure}[thbp]
\epsfxsize=7.5 cm \epsfysize=6.5cm
\epsfbox{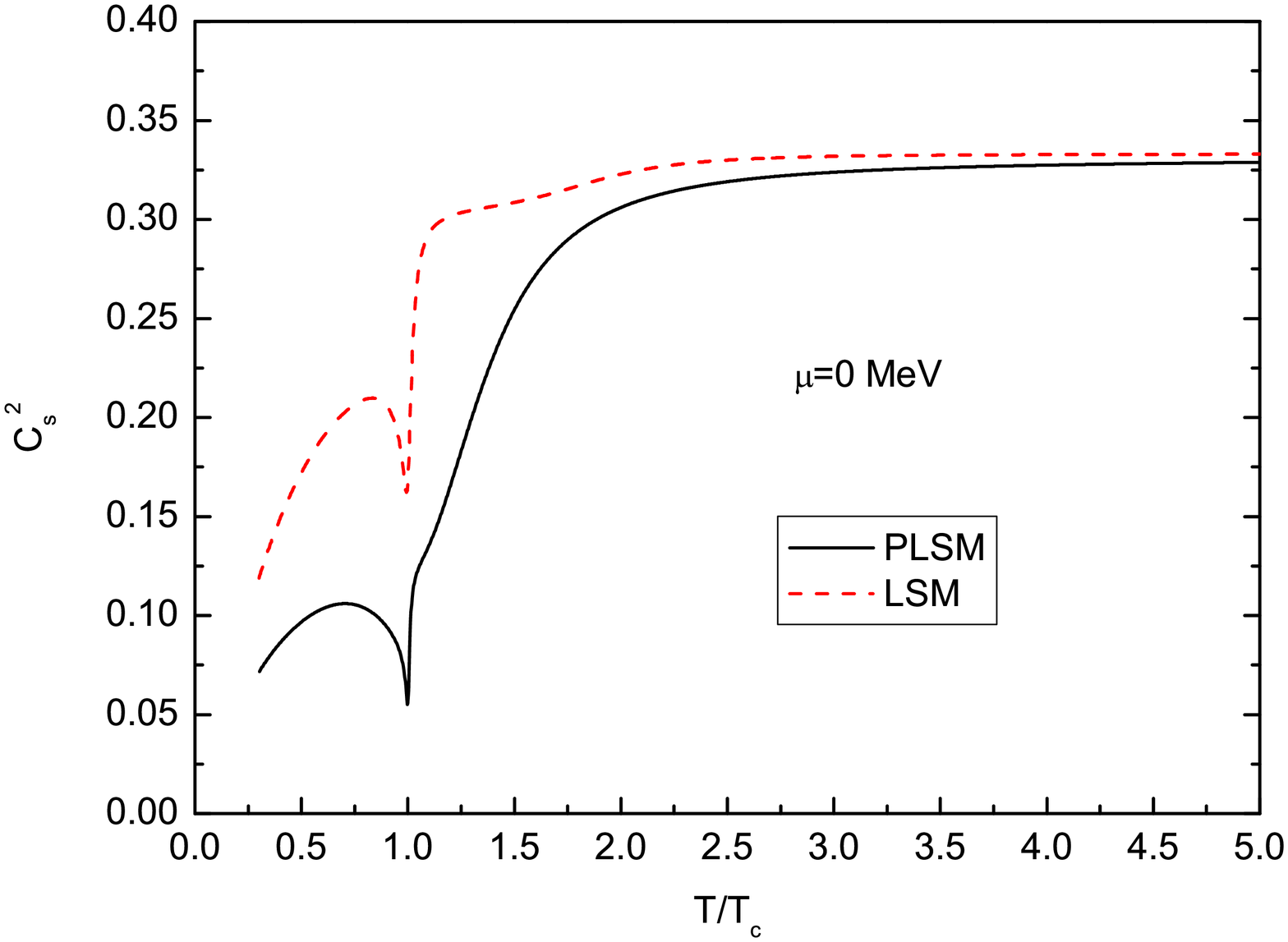}\hspace*{0.1cm} \epsfxsize=7.5 cm
\epsfysize=6.5cm \epsfbox{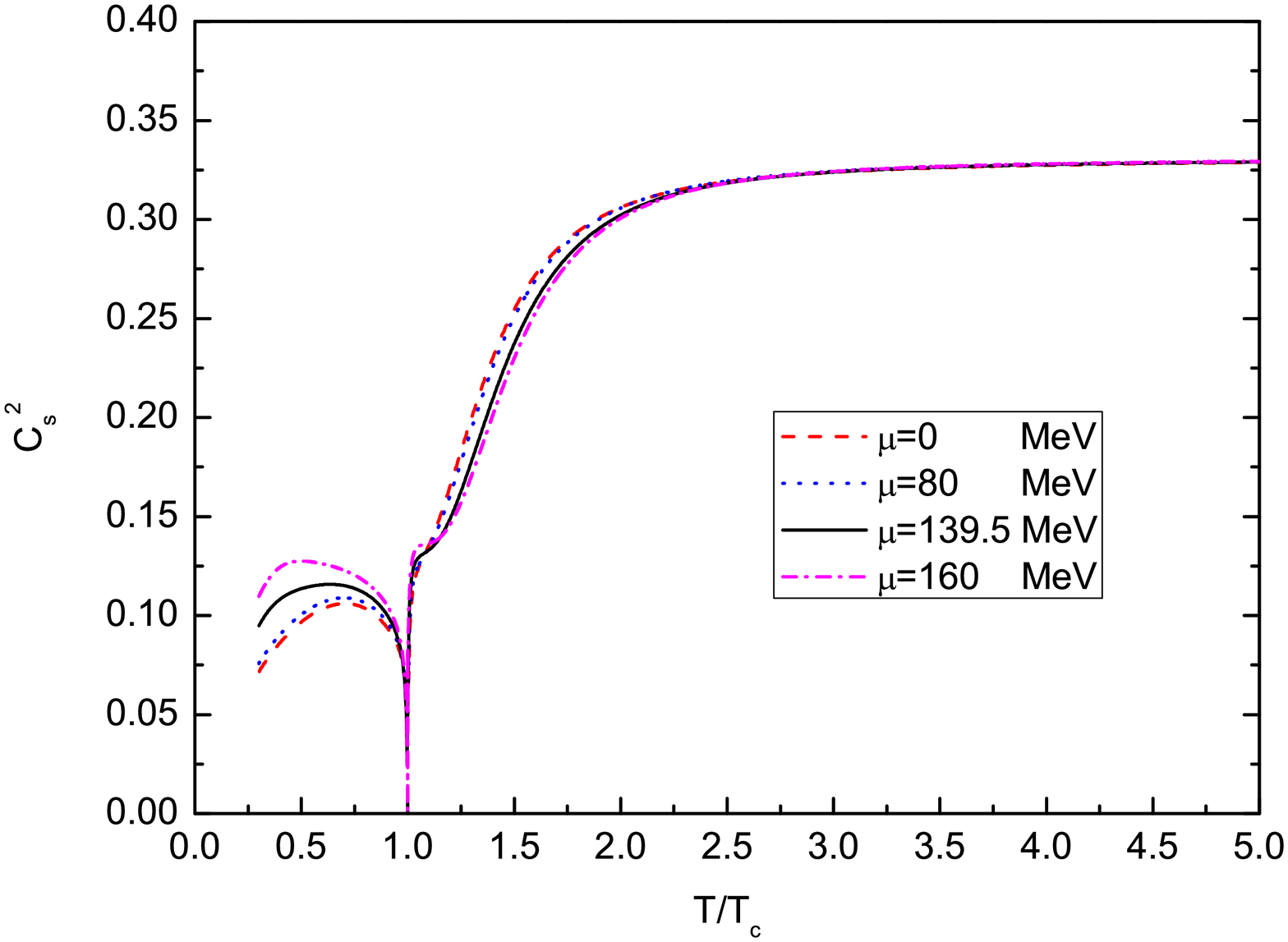}
\vskip -0.05cm \hskip 0.15 cm \textbf{( a ) } \hskip 6.5 cm \textbf{( b )} \\
 \caption{(a) The sound velocity square $C^{2}_s$ as a function of the
temperature for $\mu=0$ MeV. The solid line denotes the Polyakov
linear sigma model prediction and the dashed line denotes the linear
sigma model prediction. (b) The sound velocity square $C^{2}_s$ in
the PLSM for different chemical potentials as functions of $T/T_c$.}
\label{Fig:Fig08}
\end{figure}

In Fig.\ref{Fig:Fig04}-\ref{Fig:Fig08}, we show the trace anomaly,
energy density, pressure density, equation of state parameter,
specific heat and sound velocity square as functions of the scaled
temperature $T/T_c$.

The result of the trace anomaly of the energy-momentum tensor ${\cal
T}^{\mu\nu}$ at zero density is shown in Fig.\ref{Fig:Fig04}(a) in
comparison with lattice data for $N_{\tau}=6$ \cite{Cheng:2007jq}.
It is shown that the trace anomaly approaches the conformal value
$0$ at high temperature in both the linear sigma model and the
Polyakov linear sigma model, which agrees with the lattice result.
The trace anomaly shows a peak around $T_c$ in the linear sigma
model, and the height is only half of that of lattice result. After
including the Polyakov loop, the peak of the trace anomaly appears
at a higher temperature $1.5 ~T_c$ comparing with the lattice
result, where the peak shows up at around $1.1~T_c$. We will show
later that the appearance of the peak in the trace anomaly is not
related to the phase transition, but the change rate of the trace
anomaly is related to the phase transition.

The result of the trace anomaly of the energy-momentum tensor at
different chemical potentials as function of the scaled temperature
$T/T_c$ is shown in Fig.\ref{Fig:Fig04}(b). It is observed that the
trace anomaly changes smoothly around $T_c$ in the case of zero
chemical potential when the system experiences a crossover, but
changes sharply in the case of fist order phase transition at high
density. From Fig.\ref{Fig:Fig05}, we can see clearly that the
behavior of the trace anomaly at $T_c$ resembles that of the energy
density and the entropy density at $T_c$. We will show latter that
the change rate of the energy density and/or entropy density at
$T_c$ determine the critical behavior of other thermal quantities
like the specific heat, the sound velocity square.

Fig.\ref{Fig:Fig06} shows the pressure density over energy density
$p/\varepsilon$, which is represented in terms of equation-of-state
(EOS) parameter, at zero density and finite density, respectively.
We observe that the pressure density over energy density increases
with temperature and saturates at high temperature. Both the linear
sigma model and the Polyakov linear sigma model give very similar
results at high temperature, the pressure density over energy
density $p/\varepsilon$ saturates at a value smaller than $1/3$.
Another common feature of the $p/\varepsilon$ in the linear sigma
model and the Polyakov linear sigma model is that there is a bump
appearing at low temperature region, which is also observed in the
lattice result. Around the critical temperature $T_c$, the pressure
density over energy density $p/\varepsilon$ shows a downward cusp.
However, the minimum value of the $p/\varepsilon$ around $T_c$ is
$0.2$ in the linear sigma model, which is much larger than the
result from the Polyakov linear sigma model and the lattice QCD
data. For the Polyakov linear sigma model, the minimum of
$p/\varepsilon$ around $T_c$ is $0.075$, which is consistent with
the lattice QCD data \cite{Cheng:2007jq}. When the chemical
potential increases, from Fig.\ref{Fig:Fig06} (b), we can see that
the minimum of the $p/\varepsilon$ around $T_c$ decreases.

We show the specific heat $C_v$ as a function of the scaled
temperature $T/T_c$ in Fig.\ref{Fig:Fig07}. It it shown that there
is a sharp peak arising at $T_c$ both in the Polyakov linear sigma
model and linear sigma model. At high chemical potential when the
phase transition is of first order, the specific heat diverges. From
the definition of the specific heat $C_v=\partial
\varepsilon/\partial T$, it is easy to understand that the
appearance of the peak is due the fast change of the energy density
at the critical temperature $T_c$. The other issue presented in the
temperature behavior of $C_v$ in Fig.\ref{Fig:Fig07} is that there
is a small second peak at higher temperature, which is more visible
in the Polyakov linear sigma model than that in the linear sigma
model. This result is consistent with the PLSM model with two quark
flavors\cite{Kahara:2008yg}.

Fig.\ref{Fig:Fig08} shows the sound velocity square $C^{2}_s$ as a
function of the scaled temperature. In conformal field theories
including free field theory, the sound velocity square is always
$1/3$. In both the linear sigma model and the Polyakov linear sigma
model, the $C^{2}_s$ saturates at a value smaller than $1/3$ at high
temperature. However, near the critical temperature $T_c$, a
downward cusp shows up. At zero density, the minimum of the sound
velocity square $C^{2}_s$ at $T_c$ is around $0.05$ and $0.16$ in
the Polyakov linear sigma model and the linear sigma model,
respectively, and the value of $C^{2}_s$ at $T_c$ in the Polyakov
linear sigma model is close the lattice result. When the chemical
potential increases, the minimum of $C^{2}_s$ at $T_c$ decreases,
and it approaches $0$ in the case of first order phase transition.

\section{The bulk viscosity over entropy density ratio in the Polyakov linear sigma model}

The bulk viscosity is related to the correlation function of the
trace of the energy-momentum tensor $\theta^\mu_\mu$:
\begin{equation}
\label{kubo} \zeta = \frac{1}{9}\lim_{\omega\to
0}\frac{1}{\omega}\int_0^\infty dt \int d^3r\,e^{i\omega t}\,\langle
[\theta^\mu_\mu(x),\theta^\mu_\mu(0)]\rangle \,.
\end{equation}
According to the result derived from low energy theorem, in the low
frequency region, the bulk viscosity takes the form of
\cite{Kharzeev:2007wb}\cite{Karsch:2007jc}
\begin{eqnarray}\label{ze}
\,\zeta &=& \frac{1}{9\,\omega_0}\left\{ T^5\frac{\partial}{\partial
T}\frac{(\varepsilon-3p)}{T^4}
+16|\varepsilon_v|\right\}\,, \nonumber \\
 & = & \frac{1}{9\,\omega_0} \left\{- 16 \varepsilon+9 T S + T C_v + 16 |\varepsilon_v| \right\}\,.
\end{eqnarray}
with the negative vacuum energy density
$\varepsilon_v=\Omega_v=\Omega(\phi)|_{T=0}$, and the parameter
$\omega_0 = \omega_0(T)$ is a scale at which the perturbation theory
becomes valid. From the above formula, we can see that the bulk
viscosity is proportional to the specific heat $C_v$ near phase
transition, thus $\zeta/s$ behaves as $1/C_s^2$ near $T_c$ in this
approximation.

\begin{figure}[thbp]
\epsfxsize=7.5 cm \epsfysize=6.5cm
\epsfbox{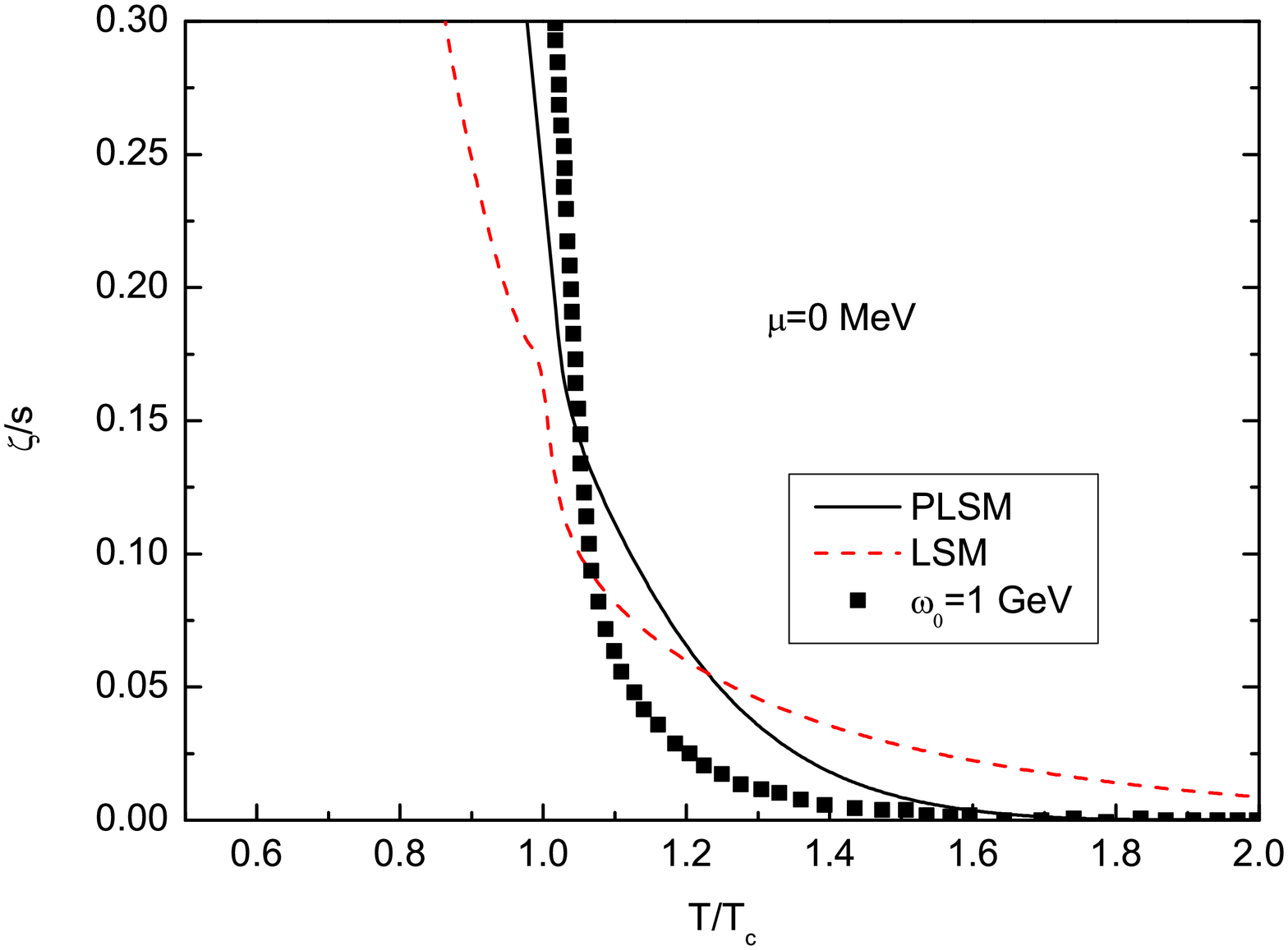}\hspace*{0.1cm} \epsfxsize=7.5 cm
\epsfysize=6.5cm \epsfbox{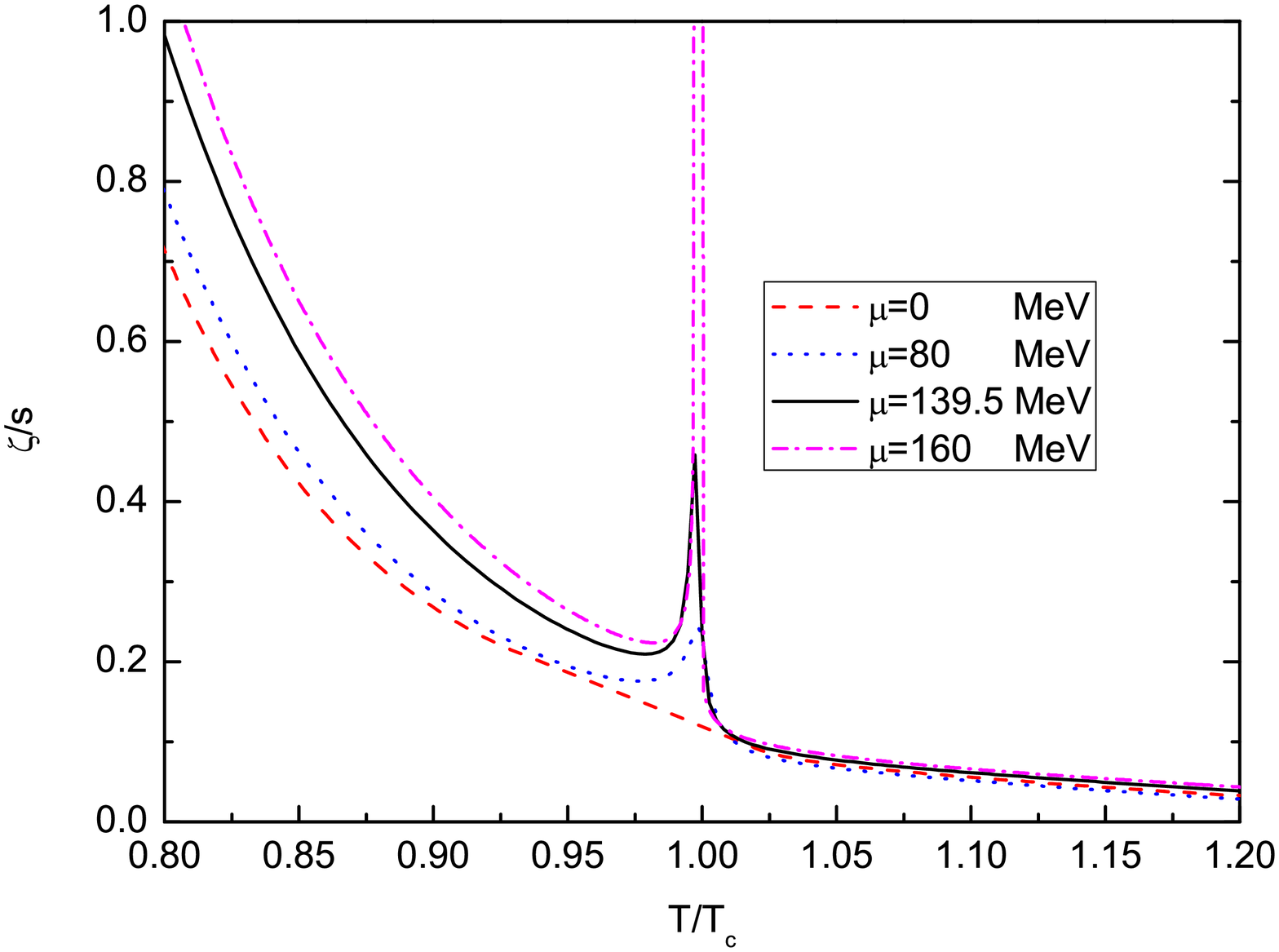}
\vskip -0.05cm \hskip 0.15 cm \textbf{( a ) } \hskip 6.5 cm \textbf{( b )} \\
 \caption{
(a) The bulk viscosity over entropy density ratio $\zeta /s$ as a
function of the temperature for $\mu=0$ MeV. The solid line denotes
the Polyakov linear sigma model prediction and the dashed line
denotes the linear sigma model prediction. Lattice data taken from
Ref.\cite{Karsch:2007jc}. (b) The bulk viscosity
over entropy density ratio $\zeta /s$ in the PLSM for different
chemical potentials as functions of $T/T_c$.} \label{Fig:Fig09}
\end{figure}

In Fig.\ref{Fig:Fig09} ~(a) and (b), we plot the bulk viscosity over
entropy density ratio $\zeta /s$ as a function of the temperature
for zero chemical potential and finite chemical potential,
respectively. From Fig.\ref{Fig:Fig09} ~(a), it is shown that, at
zero chemical potential $\mu=0$, the bulk viscosity over entropy
density $\zeta /s$ decreases monotonically with the increase of the
temperature in both the Polyakov linear sigma model and linear sigma
model, and at high temperature, $\zeta /s$ reaches its conformal
value $0$. In \cite{Karsch:2007jc}, the bulk viscosity over entropy
density of the three flavor system is extracted from lattice result,
which is shown in Fig.\ref{Fig:Fig09} ~(a) by the square. It is
observed that $\zeta /s$ in PLSM near phase transition is in very
good agreement with the lattice result in \cite{Karsch:2007jc}, i.e,
it rises sharply near phase transition. From Fig.\ref{Fig:Fig09}
~(b), which shows $\zeta /s$ as function of the scaled temperature
$T/T_c$ for different chemical potentials with $\mu=0,80,139.5, 160$
MeV, We can see that when the chemical potential increases up to
$\mu=80$ MeV, there is an upward cusp appearing in $\zeta /s$ right
at the critical temperature $T_c$. With the increase of the chemical
potential, the upward cusp becomes sharper, and the height of the
cusp increases. At the critical end point $\mu_E$ and when
$\mu>\mu_E$ for the first order phase transition, $\zeta /s$ becomes
divergent at the critical temperature.

The critical behavior of $\zeta /s$ is determined by the shape of
the trace anomaly or the energy density around the critical
temperature $T_c$. In the case of first order phase transition, the
energy density has a sudden change at the critical temperature, thus
$\zeta /s$ diverges at $T_c$. In the case of crossover, $\zeta /s$
can exhibit different behaviors near phase transition: 1) When the
energy density or entropy density changes slowly around $T_c$,
$\zeta /s$ monotonically but slowly rises up with the decrease of
temperature. 2) When the energy density or the entropy density
changes quickly near $T_c$, $\zeta /s$ monotonically but quickly
rises up when the temperature decreases. 3) When the energy density
or entropy density changes very quickly near $T_c$, an upward cusp
of $\zeta /s$ appears at $T_c$, i.e, $\zeta /s$ firstly rises
quickly when the temperature decreases to $T_c$, then jumps fast and
eventually rises up when the temperature drops further away from
$T_c$. These behaviors have been analyzed in Ref. \cite{Li:2009by}
for a toy model.

It is expected that one can distinguish whether the system
experiences a first order phase transition or a crossover from
observables which are sensitive to the bulk viscosity at RHIC
experiments. As mentioned in
Refs.\cite{Torrieri:2008ip,Torrieri:2007fb} that a sharp rise of
bulk viscosity near phase transition induces an instability in the
hydrodynamic flow of the plasma, and this mode will blow up and tear
the system into droplets. It would be interesting to investigate in
more detail how different behaviors of $\zeta/s$ affect the
observables.

However, it is noticed that the results of bulk viscosity in this
paper are based on Eq. (\ref{ze}), where the ansatz for the spectral
function \begin{equation}
\frac{\rho(\omega,\vec{0})}{\omega}=\frac{9\zeta}{\pi}
\frac{\omega_0^2}{\pi (\omega^2+\omega^2)} \label{ansatz}
\end{equation}
has been used in the small frequency, and $\omega_0$ is a scale at
which the perturbation theory becomes valid. In our calculation,
$\omega_0=10~T$, its magnitude at $T_c$ is in agreement with that
obtained in ChPT for massive pion gas system in Ref. \cite{Fraile}.
Qualitatively, the bulk viscosity corresponds to nonconformality,
thus it is reasonable to observe a sharp rising of bulk viscosity
near phase transition.  Ref. \cite{Fraile} has investigated the
correlation between the bulk viscosity and conformal breaking, and
supports the results in Ref.\cite{Kharzeev:2007wb,Karsch:2007jc}.
The sharp rising of bulk viscosity has also been observed by another
lattice result \cite{Meyer:2007dy} and in the linear sigma model
\cite{bulk-Paech-Pratt}. However, till now, no full calculation has
been done for the bulk viscosity. The frequency dependence of the
spectral density has been analyzed in Refs. \cite{Moore} and
\cite{correlation-Karsch} and the limitation of the ansatz
Eq.(\ref{ansatz}) has been discussed. From Eq. (\ref{ze}), we see
that the bulk viscosity is dominated by $C_v$ at $T_c$. If $C_v$
diverges at $T_c$, the bulk viscosity should also be divergent at
the critical point and behave as $t^{-\alpha}$. However, the
detailed analysis in the Ising model in Ref. \cite{Onuki} shows a
very different divergent behavior $\zeta \sim t^{-z\nu +\alpha}$,
with $z\simeq 3 $ the dynamic critical exponent and $\nu\simeq
0.630$ the critical exponent in the Ising system. More careful
calculation on the bulk viscosity is needed in the future.

\section{Summary and discussion}

In this paper, we have extended the linear sigma model with three
quark flavors to include certain aspects of gluon dynamics via the
Polyakov loop. The PLSM model encodes two basic features that govern
low energy QCD, spontaneous chiral symmetry breaking and
confinement, and is a framework proposed to correctly interpret
results from QCD thermodynamics and extrapolate to regions not
accessible by lattice computations.

Within the mean field approximation, we have studied the ($T,\mu$)
phase diagram of the Polyakov linear sigma model with the polynomial
form of the Polyakov loop potential. It is found that in the linear
sigma model with Polyakov loop, the three phase transitions, i.e,
the chiral restoration of $u,d$ quarks, the chiral restoration of
$s$ quark, and the deconfinement phase transition are independent
and happen at different critical temperatures. It is found that in
the PLSM, at low density, there exists two-flavor quarkyonic phase,
where the $u,d$ quarks restore chiral symmetry but still in
confinement, and at high density there exists three-flavor
quarkyonic phase where $u,d$ quarks restore chiral symmetry but
still in confinement. The linear sigma model with and without the
Polyakov loop has a general feature: there is a crossover in the low
density region and turn a first-order phase transition in the
high-density region accompanied by a critical end point (CEP). Here
we observe large discrepancies for the critical end point between
the PLSM and LSM models. For the LSM model, the critical point is
found at $(T_E,\mu_E) \simeq (92.5,216)$ MeV, while that of the PLSM
model is at $(T_E, \mu_E)\simeq (188, 139.5)$ MeV, which is close to
the lattice result $(T_E, \mu_E)=(162\pm 2~ MeV, \mu_E=120 \pm
13~MeV)$ \cite{Fodor:2004nz}. The critical chemical potential
$\mu_E$ in PLSM is much lower than that in the PNJL model with three
quark flavors where the predicted critical end point is at $\mu_E >
300$ MeV\cite{Fu:2007xc, Ciminale:2007sr}.

In order to compare our results with the lattice QCD simulations and
other models, at zero chemical potential but finite temperature, we
have investigated the thermodynamic properties and bulk viscosity in
the PLSM and LSM model. It is found that the inclusion of the
Polyakov loop is necessary in order to quantitatively fit the
lattice QCD result in the pure gauge sector. Our results in the PLSM
and LSM models show that at critical temperature $T_c$, the trace
anomaly $\Delta$, the specific heat $C_v$ show upward cusp at $T_c$.
The ratio of pressure density over energy density $p/\varepsilon$
and the square of the sound velocity $C^{2}_s$ show downward cusp at
$T_c$. These cusp behaviors at phase transition resemble lattice QCD
results. We find that the PLSM model can reproduce all the
thermodynamic and transport properties of the hot quark-gluon system
near the critical temperature.  $p/\varepsilon$ at $T_c$ is close to
the lattice QCD results $0.075$, the trace anomaly $\Delta \simeq
6.6$ at $1.3T_c$ is close to the lattice result of the peak value
$7.7$, the bulk viscosity to entropy density ratio $\zeta /s$ at
$T_c$ is around $0.2$, which is also agreement with the lattice QCD
result in Ref.\cite{Karsch:2007jc,Meyer:2007dy}.

In this work the thermodynamics of the PLSM model have been compared
to three-flavor lattice data, and it has been shown that the results
of the model indeed agree with the lattice data, but it should be
pointed that the used quark masses are different from each other,
due to the fact that the lattice data with large quark masses while
our model calculations physical values have been used. Furthermore
because the model has several adjustable parameters, the numerical
results of the critical temperature and the critical chemical
potential of the phase transitions are, of course, parameter
dependent. As above discussions for small temperatures and finite
chemical potentials the chiral phase transition is probably of
first-order while a crossover is expected at high temperatures and
small chemical potentials, this suggests the existence of at least
one critical endpoint (CEP) where the first-order transition line in
the phase diagram terminates. For the lattice simulations, although
there are much progress have been achieved in the lattice studies
for the QCD thermodynamics, at finite chemical potentials the
fermion sigma problem is still a considerable obstacle, some lattice
groups differ in their predictions. For the "CEP" problem, the
location and even the possible existence in the phase diagram is
still an open question \cite{Ejiri:2005ts}\cite{Stephanov:2007fk},
in such a manner, a comparison to the lattice data in
Ref.\cite{Fodor:2004nz} is less meaningful.

It is interesting to mention some further developments for this
work, such as extensions beyond mean field theory by using loop
expansion\cite{Herpay:2006vc}\cite{Cornwall:1974vz} or concerning both
Polyakov loop and mesonic fluctuations \cite{Rossner:2007ik}, and
consideration about matter at high baryon density for
asymmetric quark matter, and such
discussions are more relevant to the real experiment (RHIC and LHC)
or observation (neutron stars), all of above extensions leave much room
for further detailed investigations.
Especially, at high chemical potential with asymmetric quark matter
$\mu_u\neq \mu_d \neq \mu_s$, since in that case we can discuss the
pion condensate and kaon condensate in the Polyakov linear sigma
model \cite{Zhang:2006gu,Andersen:2006ys,Phat:2008zz}.
Eventually, work in this direction is in progress.

\begin{acknowledgments}
We thank S. He, B.C. Li, T. Kahara and Z. Zhang for valuable discussions. The
work of H.M. and J.J. is supported by NSFC 10904029,10905014 and the Natural Science
Foundation of Zhejiang Province under Grant No. Y7080056, Y6090345. The work
of M.H. is supported by CAS program "Outstanding young scientists
abroad brought-in", CAS key project KJCX3-SYW-N2, NSFC10735040,
NSFC10875134, and the support of K.C.Wong Education Foundation, Hong
Kong.

\end{acknowledgments}


\begin{thebibliography}{199}

\bibitem{Rischke:2003mt}
  D.~H.~Rischke,
  Prog.\ Part.\ Nucl.\ Phys.\  {\bf 52}, 197 (2004).

\bibitem{'tHooft:1976up}
  G.~'t Hooft,
  Phys.\ Rev.\ Lett.\  {\bf 37}, 8 (1976).

\bibitem{'tHooft:1976fv}
  G.~'t Hooft,
  Phys.\ Rev.\  D {\bf 14}, 3432 (1976)
  [Erratum-ibid.\  D {\bf 18}, 2199 (1978)].

\bibitem{Roder:2003uz}
  D.~Roder, J.~Ruppert and D.~H.~Rischke,
  Phys.\ Rev.\  D {\bf 68}, 016003 (2003)

\bibitem{Lenaghan:2000ey}
  J.~T.~Lenaghan, D.~H.~Rischke and J.~Schaffner-Bielich,
  Phys.\ Rev.\  D {\bf 62}, 085008 (2000).

\bibitem{Karsch:2001cy}
  F.~Karsch,
  Lect.\ Notes Phys.\  {\bf 583}, 209 (2002).


\bibitem{Polyakov:1978vu}
  A.~M.~Polyakov,
  Phys.\ Lett.\  B {\bf 72} (1978) 477.

\bibitem{Susskind:1979up}
  L.~Susskind,
  Phys.\ Rev.\  D {\bf 20}, 2610 (1979).

\bibitem{Svetitsky:1982gs}
  B.~Svetitsky and L.~G.~Yaffe,
  Nucl.\ Phys.\  B {\bf 210}, 423 (1982).

\bibitem{Svetitsky:1985ye}
  B.~Svetitsky,
  Phys.\ Rept.\  {\bf 132}, 1 (1986).

\bibitem{Fukushima:2002bk}
  K.~Fukushima,
  Annals Phys.\  {\bf 304}, 72 (2003).

\bibitem{Fukushima:2003fw}
  K.~Fukushima,
  Phys.\ Lett.\  B {\bf 591}, 277 (2004).

\bibitem{Ratti:2005jh}
  C.~Ratti, M.~A.~Thaler and W.~Weise,
  Phys.\ Rev.\  D {\bf 73}, 014019 (2006).

\bibitem{Fukushima:2008wg}
  K.~Fukushima,
  Phys.\ Rev.\  D {\bf 77}, 114028 (2008);
  [Erratum-ibid.\  D {\bf 78}, 039902 (2008)].

\bibitem{Schaefer:2007pw}
  B.~J.~Schaefer, J.~M.~Pawlowski and J.~Wambach,
  Phys.\ Rev.\  D {\bf 76}, 074023 (2007).

\bibitem{Kahara:2008yg}
  T.~Kahara and K.~Tuominen,
  Phys.\ Rev.\  D {\bf 78}, 034015 (2008).

\bibitem{Schaefer:2008ax}
  B.~J.~Schaefer and M.~Wagner,
  arXiv:0812.2855 [hep-ph].

\bibitem{Megias:2004hj}
  E.~Megias, E.~Ruiz Arriola and L.~L.~Salcedo,
  Phys.\ Rev.\  D {\bf 74}, 065005 (2006).

\bibitem{Ghosh:2006qh}
  S.~K.~Ghosh, T.~K.~Mukherjee, M.~G.~Mustafa and R.~Ray,
  Phys.\ Rev.\  D {\bf 73}, 114007 (2006).

\bibitem{Ratti:2006wg}
  C.~Ratti, S.~Roessner, M.~A.~Thaler and W.~Weise,
  Eur.\ Phys.\ J.\  C {\bf 49}, 213 (2007).

\bibitem{Mukherjee:2006hq}
  S.~Mukherjee, M.~G.~Mustafa and R.~Ray,
  Phys.\ Rev.\  D {\bf 75}, 094015 (2007).

\bibitem{Roessner:2006xn}
  S.~Roessner, C.~Ratti and W.~Weise,
  Phys.\ Rev.\  D {\bf 75}, 034007 (2007).

\bibitem{Fu:2007xc}
  W.~j.~Fu, Z.~Zhang and Y.~x.~Liu,
  Phys.\ Rev.\  D {\bf 77}, 014006 (2008).

\bibitem{Ciminale:2007sr}
  M.~Ciminale, R.~Gatto, N.~D.~Ippolito, G.~Nardulli and M.~Ruggieri,
  Phys.\ Rev.\  D {\bf 77}, 054023 (2008).

\bibitem{Abuki:2008nm}
  H.~Abuki, R.~Anglani, R.~Gatto, G.~Nardulli and M.~Ruggieri,
  Phys.\ Rev.\  D {\bf 78}, 034034 (2008).

\bibitem{GomezDumm:2008sk}
  D.~Gomez Dumm, D.~B.~Blaschke, A.~G.~Grunfeld and N.~N.~Scoccola,
  Phys.\ Rev.\  D {\bf 78}, 114021 (2008).

\bibitem{Fu:2009zs}
  W.~j.~Fu and Y.~x.~Liu,
  Phys.\ Rev.\  D {\bf 79}, 074011 (2009).

\bibitem{Zhang:2006gu}
  Z.~Zhang and Y.~X.~Liu,
  Phys.\ Rev.\  C {\bf 75}, 064910 (2007).

\bibitem{Sasaki:2006ww}
  C.~Sasaki, B.~Friman and K.~Redlich,
  Phys.\ Rev.\  D {\bf 75}, 074013 (2007).

\bibitem{Ratti:2007jf}
  C.~Ratti, S.~Roessner and W.~Weise,
  Phys.\ Lett.\  B {\bf 649}, 57 (2007).

\bibitem{Herpay:2006vc}
  T.~Herpay and Z.~Szep,
  Phys.\ Rev.\  D {\bf 74}, 025008 (2006).

\bibitem{Cornwall:1974vz}
  J.~M.~Cornwall, R.~Jackiw and E.~Tomboulis,
  Phys.\ Rev.\  D {\bf 10}, 2428 (1974).

\bibitem{Cheng:2007jq}
  M.~Cheng {\it et al.},
  Phys.\ Rev.\  D {\bf 77}, 014511 (2008).




\bibitem{Schaefer:2008hk}
  B.~J.~Schaefer and M.~Wagner,
  Phys.\ Rev.\  D {\bf 79}, 014018 (2009).

\bibitem{Lenaghan:2000kr}
  J.~T.~Lenaghan,
  Phys.\ Rev.\  D {\bf 63}, 037901 (2001).

\bibitem{Scavenius:2000qd}
  O.~Scavenius, A.~Mocsy, I.~N.~Mishustin and D.~H.~Rischke,
  Phys.\ Rev.\  C {\bf 64}, 045202 (2001)

\bibitem{Kapusta:2006pm}
  J.~I.~Kapusta and C.~Gale,
  ``\textit{Finite-temperature field theory: Principles and applications},''
( Cambridge University Press, UK, 2006).

\bibitem{Fukushima:2006uv}
  K.~Fukushima and Y.~Hidaka,
  Phys.\ Rev.\  D {\bf 75}, 036002 (2007)
  [arXiv:hep-ph/0610323].

\bibitem{Abuki:2009dt}
  H.~Abuki and K.~Fukushima,
  arXiv:0901.4821 [hep-ph].

\bibitem{Fodor:2004nz}
  Z.~Fodor and S.~D.~Katz,
  JHEP {\bf 0404}, 050 (2004)
  [arXiv:hep-lat/0402006].

\bibitem{McLerran:2007qj}
  L.~McLerran and R.~D.~Pisarski,
  Nucl.\ Phys.\  A {\bf 796}, 83 (2007)
  [arXiv:0706.2191 [hep-ph]].

\bibitem{Kharzeev:2007wb}
  D.~Kharzeev and K.~Tuchin,
  JHEP {\bf 0809}, 093 (2008).

\bibitem{Karsch:2007jc}
  F.~Karsch, D.~Kharzeev and K.~Tuchin,
  Phys.\ Lett.\  B {\bf 663}, 217 (2008).

\bibitem{Li:2009by}
  B.~C.~Li and M.~Huang,
  arXiv:0903.3650 [hep-ph], and references therein.

\bibitem{Torrieri:2008ip}
  G.~Torrieri and I.~Mishustin,
  Phys.\ Rev.\  C {\bf 78}, 021901 (2008).

\bibitem{Torrieri:2007fb}
  G.~Torrieri, B.~Tomasik and I.~Mishustin,
  Phys.\ Rev.\  C {\bf 77}, 034903 (2008).

\bibitem{Fraile}
D.~Fernandez-Fraile and A.~G.~Nicola,
  Phys.\ Rev.\ Lett.\  {\bf 102}, 121601 (2009)
  [arXiv:0809.4663 [hep-ph]],
  D.~Fernandez-Fraile and A.~Gomez Nicola,
  Eur.\ Phys.\ J.\  C {\bf 62}, 37 (2009)
  [arXiv:0902.4829 [hep-ph]].

\bibitem{Meyer:2007dy}
  H.~B.~Meyer,
  Phys.\ Rev.\ Lett.\  {\bf 100}, 162001 (2008).

\bibitem{bulk-Paech-Pratt} K.~Paech and S.~Pratt,
  Phys.\ Rev.\  C {\bf 74}, 014901 (2006).

\bibitem{Moore}
  G.~D.~Moore and O.~Saremi,
  JHEP {\bf 0809}, 015 (2008)

\bibitem{correlation-Karsch}
  K.~Huebner, F.~Karsch and C.~Pica,
  arXiv:0808.1127 [hep-lat].

\bibitem{Onuki}
 A.~Onuki,
Phys. \ Rev. \ E {\bf 55}, 403 (1997).

\bibitem{Ejiri:2005ts}
  S.~Ejiri,
  Phys.\ Rev.\  D {\bf 73}, 054502 (2006).

\bibitem{Stephanov:2007fk}
  M.~A.~Stephanov,
  PoS {\bf LAT2006}, 024 (2006).


\bibitem{Rossner:2007ik}
  S.~Roessner, T.~Hell, C.~Ratti and W.~Weise,
  Nucl.\ Phys.\  A {\bf 814}, 118 (2008).

\bibitem{Andersen:2006ys}
  J.~O.~Andersen,
  Phys.\ Rev.\  D {\bf 75}, 065011 (2007).

\bibitem{Phat:2008zz}
  T.~H.~Phat, N.~V.~Long, N.~T.~Anh and L.~V.~Hoa,
  Phys.\ Rev.\  D {\bf 78}, 105016 (2008).

\end{thebibliography}
\end{document}